\documentclass[twocolumn]{aastex62}
\usepackage{hyperref,graphicx,wrapfig,booktabs,amsmath,comment,placeins}
\graphicspath{{images/}}
\usepackage{color,soul}

\def\gridline#1{\vskip6pt\hbox to\hsize{#1}\vskip6pt}

\begin{document}

\title{A Catalog of 220 Offset and Dual AGNs: Increased AGN Activation in Major Mergers and Separations under 4 kpc}

\author{Aaron Stemo}
\affiliation{Department of Astrophysical and Planetary Sciences, University of Colorado, Boulder, CO 80309, USA}

\author{Julia M. Comerford}
\affiliation{Department of Astrophysical and Planetary Sciences, University of Colorado, Boulder, CO 80309, USA}

\author{R. Scott Barrows}
\affiliation{Department of Astrophysical and Planetary Sciences, University of Colorado, Boulder, CO 80309, USA}

\author{Daniel Stern}
\affiliation{Jet Propulsion Laboratory, California Institute of Technology, 4800 Oak Grove Drive, Pasadena, CA 91109, USA}

\author{Roberto J. Assef}
\affiliation{N\'{u}cleo de Astronom\'{\i}a de la Facultad de Ingenier\'{\i}a y Ciencias, Universidad Diego Portales, Av. Ej\'{e}rcito Libertador 441, Santiago, Chile}

\author{Roger L. Griffith}
\affiliation{Jet Propulsion Laboratory, California Institute of Technology, 4800 Oak Grove Drive, Pasadena, CA 91109, USA}

\author{Aimee Schechter}
\affiliation{Department of Astrophysical and Planetary Sciences, University of Colorado, Boulder, CO 80309, USA}

\begin{abstract}

During galaxy mergers, gas and dust is driven towards the centers of merging galaxies, triggering enhanced star formation and supermassive black hole (SMBH) growth. Theory predicts that this heightened activity peaks at SMBH separations $<$20 kpc; if sufficient material accretes onto one or both of the SMBHs for them to become observable as active galactic nuclei (AGNs) during this phase, they are known as offset and dual AGNs, respectively. To better study these systems, we have built the ACS-AGN Merger Catalog, a large catalog ($N=220$) of uniformly selected offset and dual AGN observed by \textit{HST} at $0.2<z<2.5$ with separations $<$20 kpc. Using this catalog, we answer many questions regarding SMBH -- galaxy coevolution during mergers. First, we confirm predictions that the AGN fraction peaks at SMBH pair separations $<$10 kpc; specifically, we find that the fraction increases significantly at pair separations of $<$4 kpc. Second, we find that AGNs in mergers are preferentially found in major mergers and that the fraction of AGNs found in mergers follows a logarithmic relation, decreasing as merger mass ratio increases. Third, we do not find that mergers (nor the major or minor merger subpopulations) trigger the most luminous AGNs. Finally, we find that nuclear column density, AGN luminosity, and host galaxy star formation rate have no dependence on SMBH pair separation or merger mass ratio in these systems, nor do the distributions of these values differ significantly from that of the overall AGN population.

\end{abstract}

\keywords{Galaxy mergers (608); AGN host galaxies (2017); Active galactic nuclei (16); Galaxy evolution (594); Galaxy classification systems (582)}

\section{Introduction}\label{sec: intro}

Mergers between two galaxies, each with a supermassive black hole (SMBH) at its center, result in a pair of SMBHs in the merger remnant. The paired SMBHs slowly spiral toward the center of mass of the newly merged system and remain in a $<$20 kpc separation ``dual" phase for $\sim$100 Myr \citep{Begelman1980,Milosavljevic2001}, before dynamical friction drives them into a sub-pc gravitationally bound binary and they eventually coalesce. During this dual phase, significant gas can be driven inward towards the center of the merger remnant and onto the SMBHs, fueling their growth, and making them observable as active galactic nuclei (AGNs); this should also enhance star formation in the host galaxy \citep[e.g.][]{Joseph1985,Hopkins2008,Hopkins2009,Knapen2015}

When one or both of the SMBHs are accreting in this dual phase, they are known as offset AGNs \citep[e.g.][]{Comerford2014a} or dual AGNs \citep[e.g.][]{Gerke2007,Comerford2009}, respectively. During this phase, the outer stars are tidally stripped away, but the merging galaxies retain their central stellar bulges \citep{Liu2010,Fu2011,Rosario2011}; these stellar bulges contain the SMBHs. Therefore, offset AGNs are merging galaxy systems where one stellar bulge hosts an AGN and one does not; presumably the latter hosts a quiescent SMBH. Similarly, dual AGNs are merging galaxy systems in which both stellar bulges host an AGN.

There are strong correlations between observed properties of the SMBH and host galaxy properties, such as the M-$\sigma$ relation \citep[e.g.][]{Magorrian1998,Ferrarese2000,Gebhardt2000,Greene2006}, which seems to indicate a connection between SMBH growth and the evolution of its host galaxy. In order to understand this connection, studies of systems hosting AGNs during periods of growth are necessary. Offset and dual AGNs are powerful observational tools for studies of SMBH and galaxy co-evolution as they are direct probes of the state of a SMBH and its host galaxy during a merger event. However, their utility has been limited because the number of known systems is small.
 
 A large sample of offset and dual AGNs could address open questions in the field of SMBH and galaxy co-evolution. For example, some simulations predict that major mergers trigger the most luminous AGNs \citep{Hopkins2009}. However, observational studies have found conflicting results about whether the most luminous AGNs are preferentially triggered in mergers, primarily due to the lack of a large, clean sample of AGNs in merging galaxy systems \citep[e.g.][]{Kocevski2012,Treister2012,Villforth2014,Villforth2017}. Other simulations predict that the peak of SMBH growth in mergers occurs when the paired SMBHs are separated by 1 -- 10 kpc \citep{VanWassenhove2012} or 0.1 -- 2 kpc \citep{Blecha2013}. Observations have not yet been able to test these predictions due to the limited number of known systems with SMBH pair separations $<$10 kpc, but they do verify the trend in the 10 -- 100 kpc range \citep{Ellison2011,Koss2012}. Lastly, a large sample of these systems could test whether findings of increased star formation in mergers \citep[e.g.][]{Cox2008,Ellison2008,Patton2013a} also apply to merging systems hosting AGN at small separations.

Since offset and dual AGNs are a promising avenue of approach to study SMBH and galaxy co-evolution, there have been many searches recently to find them. While most initial findings of these systems were serendipitous \citep[e.g.][]{Komossa2003,Bianchi2008}, recent studies have attempted a more systematic approach. One such method is looking for the spectroscopic signatures of the AGNs in these systems, including double-peaked emission lines or single-peaked emission lines with a velocity offset relative to the host galaxy. While some of the galaxies identified in these studies have been confirmed as offset and dual AGNs \citep{Fu2011,Liu2013,Muller-Sanchez2015,Comerford2015}, many have not; this is because in most galaxies, AGN outflows and gas kinematics produce velocity shifted and double-peaked emission lines that mimic the signatures of offset and dual AGNs \citep{Rosario2010, Shen2011, Comerford2012, Fu2012, Barrows2013, Mcgurk2015, Nevin2016}. Work by \cite{Barrows2016} found offset and dual AGNs by searching for X-ray sources offset from their optical counterpart host galaxies, but the number able to be found using this method was limited due to the spatial resolution limits of the X-ray and optical observations. Still other work has focused on using surveys to search for morphological signatures of mergers, followed by spectroscopy, radio, or X-ray observations to verify their sample \citep{Koss2012, Fu2015, Satyapal2017, Silverman2020}. This accentuates a problem in finding these systems --- observations of offset and dual AGNs are difficult because of the high spatial resolution imaging and/or spectroscopy needed to differentiate the stellar bulges associated with a SMBH pair at these separations; this is also why currently known offset and dual AGNs are mostly limited to the low-redshift Universe ($z\lesssim0.2$).

The Advanced Camera for Surveys (ACS) on the \textit{Hubble Space Telescope} (\textit{HST}) has an angular resolution of 0\farcs05. This makes it ideal for observations requiring high resolution, such as detecting the multiple stellar bulges associated with offset and dual AGNs. By choosing \textit{HST} galaxies in deep survey fields, such as GEMS, COSMOS, GOODS, and AEGIS, multi-wavelength data can be used to select AGNs, while galaxy morphological fitting can select galaxy mergers. This provides an approach for detecting offset and dual AGNs in greater numbers and at higher redshifts than ever before, enabling them to be used for statistical studies of AGN activation and SMBH -- galaxy coevolution for the first time.

Here we present a catalog and analysis of 220 offset and dual AGNs, identified using a new systematic method for finding offset and dual AGNs in large, multi-wavelength \textit{HST} galaxy surveys. Section \ref{sec: ACS-AGN} discusses our initial sample of galaxies (the ACS-AGN Catalog). In Section \ref{sec:Bulge Select} we present the methods by which we modeled and selected our offset and dual AGN sample from the ACS-AGN Catalog, while Section \ref{sec: bias} discusses the biases in our methods and how we corrected for them. Section \ref{sec: gal props} discusses the AGN and host galaxy properties of our offset and dual AGNs, both in comparison to a general AGN population and in relation to their merger parameters, and Section \ref{sec: results} presents our findings on AGN triggering in mergers. Finally, in Section \ref{sec: conclusions} we discuss our conclusions. Throughout this paper, we use the Planck 2015 cosmology of $H_{0} = 67.8$ km s$^{-1}$ Mpc$^{-1}$, $\Omega_{M}=0.308$, and $\Omega_{\Lambda}=0.692$ \citep{Planck2016}.

\begin{figure*}[t]
	\centering
	\includegraphics[width=0.98\textwidth]{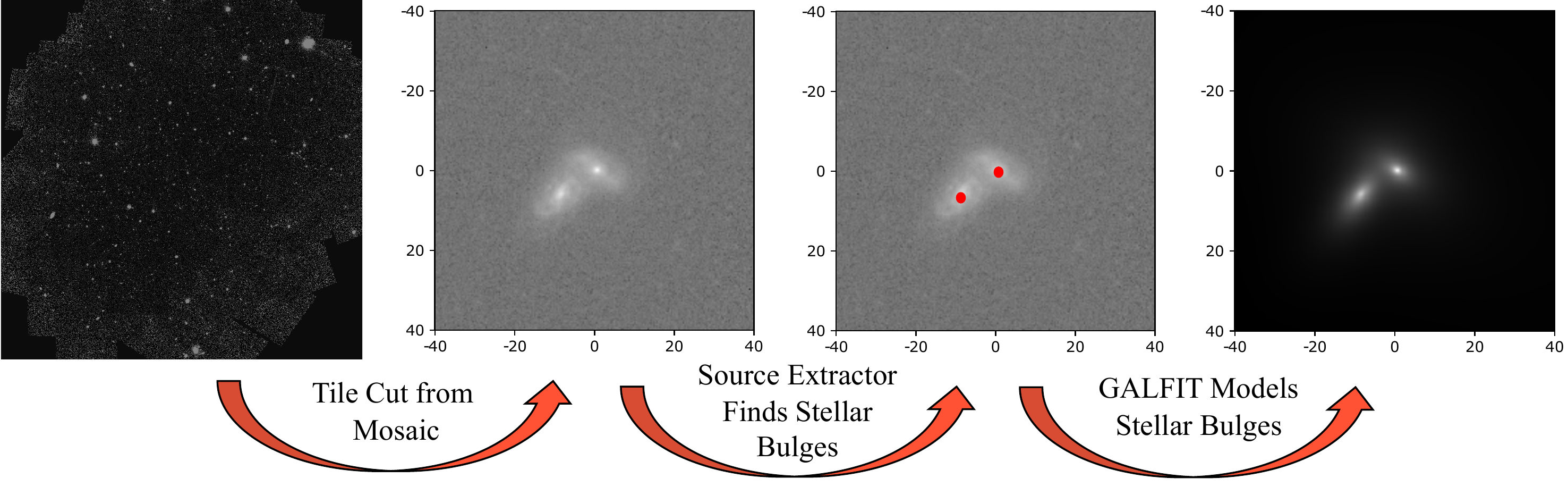}
	\vspace{-6pt}	
	\caption{A flowchart summarizing the steps that we use to identify and model offset and dual AGNs. The axes of the tiles are in units of kpc, with the origin centered on the AGN host galaxy. We purposefully choose a simple galaxy system to show here as an example; a typical galaxy tile would have more sources found by Source Extractor, some of which may be removed after they fail to meet our modeling criteria, and there would be many iterations of GALFIT modeling (Section \ref{subsec: fitting}). The system shown here has a merger mass ratio of $\sim$4 and a bulge separation of $\sim$11 kpc.}
	\label{plot: method}
	\vspace{6pt}
\end{figure*}

\section{Initial Galaxy Sample} \label{sec: ACS-AGN}

Our parent galaxy sample was the Advanced Camera for Surveys Active Galactic Nuclei Catalog \citep[ACS-AGN;][hereafter referred to as S20]{Stemo2020}. The ACS-AGN is a catalog of 2585 AGN host galaxies that were imaged with \textit{HST}/ACS and span a redshift range of $0.2<z<2.5$. These AGN host galaxies are located in four deep survey fields: the Galaxy Evolution from Morphologies and SEDs (GEMS) survey \citep{Caldwell2008}, the Cosmological Evolutionary Survey (COSMOS) \citep{Scoville2007}, the Great Observatories Origins Deep Survey (GOODS) \citep{Dickinson2003}, and the All- wavelength Extended Groth Strip International Survey (AEGIS) \citep{Davis2007}. 

These AGNs were selected by applying mid-infrared and X-ray AGN selection criteria to \textit{Spitzer} and \textit{Chandra} data available in these survey fields. The mid-infrared color cut described by \cite{Donley2012} was used for \textit{Spitzer} observed galaxies and a rest-frame X-ray luminosity cut in the 2--10 keV band of $L_{2-10}>10^{42}$ erg s$^{-1}$ was used for \textit{Chandra} observed galaxies. Of the 2585 AGN host galaxies in the ACS-AGN Catalog, 1065 are infrared selected and 1945 are X-ray selected.

While redshift data was already available for these galaxies \citep{Griffith2012}, AGN and host galaxy properties were not. These properties were calculated from SEDs created for each galaxy from multi-wavelength photometric data using the SED template fitting software \texttt{LRT} \citep{Assef2010}. Specifically, AGN bolometric luminosity ($L_{AGN}$) was calculated by integrating the AGN component of the SED model. Stellar mass ($M_{*}$) was calculated using the relation between the SED modeled (excluding the AGN component) $g'-r'$ color and $M/L_r$ from \cite{Bell2003}. Host galaxy star formation rate (SFR) was calculated from the SED modeled (excluding the AGN component) 2800 \AA$\,$ monochromatic luminosity as described in \cite{Madau1998}. Finally, the nuclear neutral hydrogen column density ($N_{H}$) was calculated from the SED modeled extinction value ($E_{B-V}$) using a conversion factor derived from \cite{Maiolino2001} and \cite{Burtscher2016}.

The parent galaxy sample has a median redshift of $\langle z \rangle \approx$ 1.15, and contains AGN host galaxies that cover a significant parameter space: $L_{AGN}$ [erg s$^{-1}$ cm$^{2}$] has a range of $10^{42} \lesssim L_{AGN} \lesssim 10^{47}$, with a median of $10^{44.7}$; $M_{*}$ [$M_{\odot}$] has a range of $10^{9} \lesssim M_{*} \lesssim 10^{12}$, with a median of $10^{10.6}$; SFR [$M_{\odot}$ yr$^{-1}$] has a range of $10^{-1} \lesssim$ SFR $\lesssim 10^{2}$, with a median of $10^{0.55}$; while $N_{H}$ [cm$^{-2}$] has a range of $10^{20.5} \lesssim N_{H} \lesssim 10^{23}$, with a median of $10^{21.4}$. This sample is found to lie generally below the star-forming main sequence and also shows correlated behavior between SFR and $L_{AGN}$, most likely due to a mutual dependence on galaxy mass.

The AGN selection process, SED creation, AGN and host galaxy property calculation, and analysis is described in more detail in S20.

\section{Galaxy Modeling \& \\ Merger Identification} \label{sec:Bulge Select}

The utility of computers in identifying, classifying, and decomposing galaxies in astronomical images has been growing as the effectiveness of image analysis software has increased. We use two such software packages, \texttt{Source Extractor} \citep{Bertin1996}  and \texttt{GALFIT} \citep{Peng2002}, to identify and model multiple stellar bulge components in \textit{HST} images of the ACS-AGN galaxies in order to identify offset and dual AGN candidates.

In Section \ref{subsec:tiles}, we create tiles of the ACS-AGN galaxies. We then use \texttt{Source Extractor} and \texttt{GALFIT} to model the morphology of the ACS-AGN galaxies in Section \ref{subsec: fitting}. Finally, in Section \ref{subsec: false pos}, we eliminate false positives and select our offset and dual AGN sample. 

\begin{figure*}[t]
	\centering
	\includegraphics[width=0.98\textwidth]{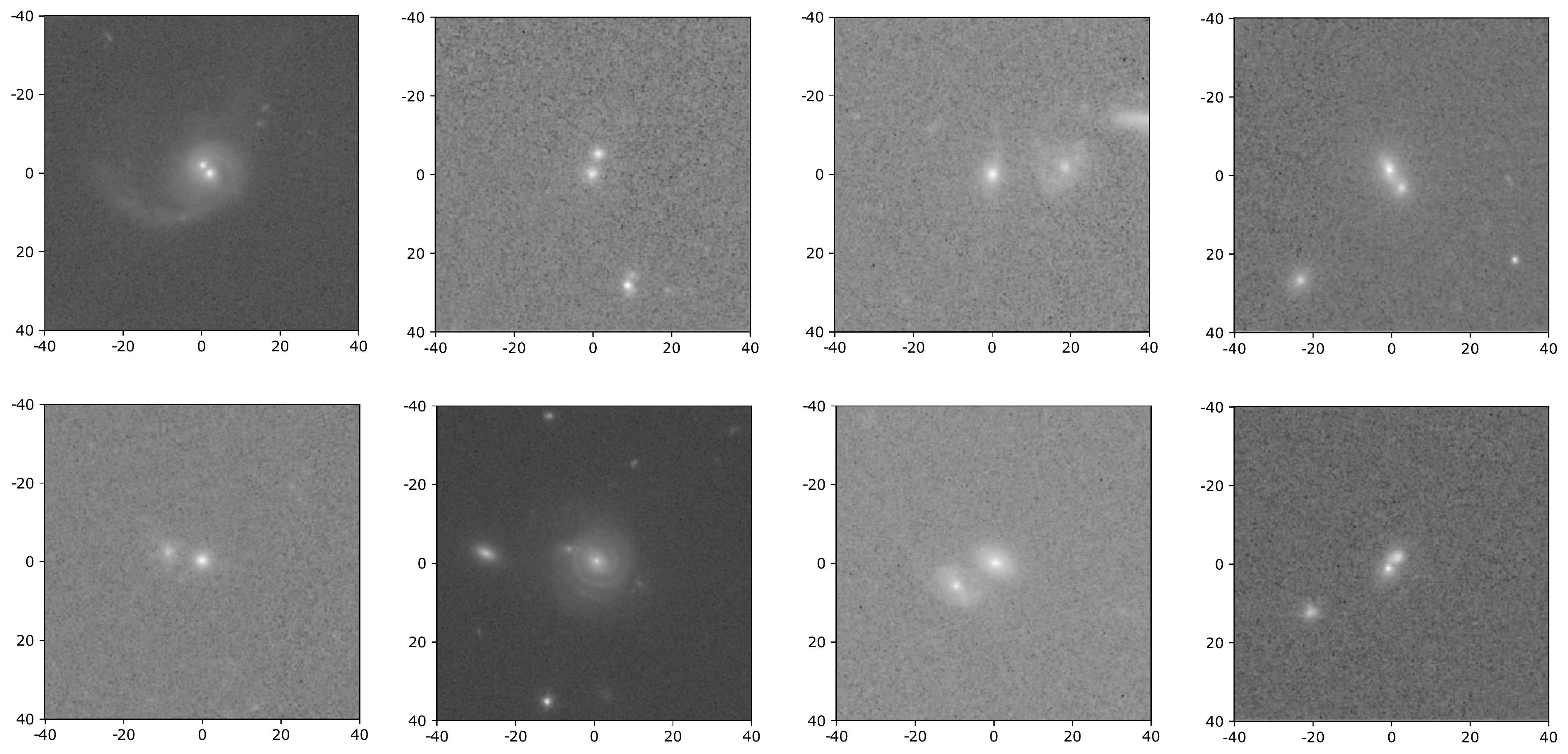}
	\vspace{-6pt}	
	\caption{Examples of offset and dual AGNs identified and included in the ACS-AGN Merger Catalog. The axes of the tiles are in units of kpc, with the origin centered on the AGN host galaxy.}
	\label{plot: examples}
	\vspace{6pt}
\end{figure*}

\subsection{Galaxy Tiling} \label{subsec:tiles}

In order to best select stellar bulges associated with SMBHs, we chose to examine them in the F814W filter of \textit{HST}/ACS --- the reddest available ACS filter that has been used to observe the entire ACS-AGN sample. This band peaks at 8333 \AA, with a FWHM of 2511 \AA, and is excellent for tracing a galaxy's stellar component in our sample's redshift range. Specifically, the F814W band outperforms available shorter wavelength bands, such as the F438W and F606W, at this task because those bands tend to be dominated by light from ionized gas, not stellar emission, which peaks further into the infrared \citep[e.g.][]{Comerford2017}. 

Of the four surveys from which our sample is drawn, only the AEGIS and COSMOS surveys have observed and provide mosaics of their respective fields in the F814W filter. Fortunately, the CANDELS survey \citep{Grogin2011} observed the remaining fields (GEMS and GOODS) in the F814W and provide mosaics of these observations. It also observed parts of the COSMOS and AEGIS fields, but not in their entirety (for further detail, see \citealt{Grogin2011} for the relevant observation maps). Therefore, where multiple surveys' observations exist for our sample, the CANDELS mosaics are the deepest available, but only marginally so --- where they are available, CANDELS Wide observations increase the depth in the COSMOS field from a 5$\sigma$ AB magnitude of 27.2 to 27.7 and from a 5$\sigma$ AB magnitude of 28.1 to 28.2 in the AEGIS field. The F814W 5$\sigma$ depth of the CANDELS Deep + Wide observations in the GEMS and GOODS fields are AB magnitude 28.8.

We gathered \textit{HST}/ACS publicly available F814W mosaics from the COSMOS \citep{Scoville2007}, AEGIS \citep{Davis2007}, and CANDELS \citep{Grogin2011} surveys. These mosaics are science products that have undergone significant processing, including calibration (e.g., bias and dark subtraction, gain correction, flat-fielding, bad pixel rejection, and low-level background removal), astrometric registration, cosmic ray cleaning, and have been co-added with MultiDrizzle  \citep{Koekemoer2007}. 

From these mosaics we created 80 kpc $\times$ 80 kpc tiles centered on the ACS-AGN galaxies. This size is significantly larger than the maximum 20 kpc separation radius we are allowing for our offset and dual AGN candidates, but allows for more accurate background estimation and subsequent fitting by \texttt{GALFIT}. For our galaxies that were observed in the F814W by CANDELS, we created tiles from the CANDELS mosaics; for those not within the boundaries of the CANDELS mosaics, we created tiles from the original survey mosaics in the F814W.

Of the 2585 ACS-AGN galaxies, we could not create tiles for 19 of them. These galaxies were in a region of the AEGIS field where no F814W mosaics were available at the time of publication. This left 2566 active galaxies that we created tiles for and could be analyzed for multiple stellar bulges.

\begin{deluxetable*}{lllll}[t!]
	\tablewidth{0pt}
	\tablecolumns{3}
	\tablecaption{Data Fields in the ACS-AGN Merger Catalog\label{tab: catalog}}
	\tablehead{
		\colhead{No. $\;\;\;$}&
		\colhead{Field $\;\;\;\;\;\;\;\;\;\;\;\;\;\;\;\;\;\;\;\;$} &
		\colhead{Note}
	}
	\startdata
	1 & ID & ACS-AGN Catalog specific unique identifier \\
	2 & RA & right ascension [J2000, decimal degrees] \\
	3 & DEC & declination [J2000, decimal degrees] \\
	4 & Z & redshift used \\
	5 & SPECZ & spectroscopic redshift \\
	6 & PHOTOZ & photometric redshift \\
	7 & Spitzer{\_}AGN & if AGN was selected in \textit{Spitzer} [Boolean] \\
	8 & Chandra{\_}AGN & if AGN was selected in \textit{Chandra} [Boolean] \\
	9 & L{\_}bol{\_}sed{\_}md & AGN bolometric luminosity calculated from SED, median [erg s$^{-1}$] \\
	10 & L{\_}bol{\_}sed{\_}lo & AGN bolometric luminosity calculated from SED, lower bound [erg s$^{-1}$] \\
	11 & L{\_}bol{\_}sed{\_}hi & AGN bolometric luminosity calculated from SED, upper bound [erg s$^{-1}$] \\
	12 & L{\_}x{\_}md & 2 -- 10 keV restframe luminosity, median [erg s$^{-1}$] \\
	13 & L{\_}x{\_}lo & 2 -- 10 keV restframe luminosity, lower bound [erg s$^{-1}$] \\
	14 & L{\_}x{\_}hi & 2 -- 10 keV restframe luminosity, upper bound [erg s$^{-1}$] \\
	15 & L{\_}bol{\_}x{\_}md & AGN bolometric luminosity calculated from X-ray, median [erg s$^{-1}$] \\
	16 & L{\_}bol{\_}x{\_}lo & AGN bolometric luminosity calculated from X-ray, lower bound [erg s$^{-1}$] \\
	17 & L{\_}bol{\_}x{\_}hi & AGN bolometric luminosity calculated from X-ray, upper bound [erg s$^{-1}$] \\
	18 & M{\_}star{\_}md & galaxy stellar mass, median [$M_{\odot}$] \\
	19 & M{\_}star{\_}lo & galaxy stellar mass, lower bound [$M_{\odot}$] \\
	20 & M{\_}star{\_}hi & galaxy stellar mass, upper bound [$M_{\odot}$] \\
	21 & SFR{\_}md & star formation rate, median [$M_{\odot}$ yr$^{-1}$] \\
	22 & SFR{\_}lo & star formation rate, lower bound [$M_{\odot}$ yr$^{-1}$] \\
	23 & SFR{\_}hi & star formation rate, upper bound [$M_{\odot}$ yr$^{-1}$] \\
	24 & Nh{\_}md & nuclear column density, median [cm$^{-2}$] \\
	25 & Nh{\_}lo & nuclear column density, lower bound [cm$^{-2}$] \\
	26 & Nh{\_}hi & nuclear column density, upper bound [cm$^{-2}$] \\
	27 & SFR{\_}norm{\_}md & normalized star formation rate, median \\
	28 & SFR{\_}norm{\_}lo & normalized star formation rate, lower bound \\
	29 & SFR{\_}norm{\_}hi & normalized star formation rate, upper bound \\
	30 & Sep{\_}12 & separation between primary (active) and secondary components [kpc] \\
	31 & Ratio{\_}12 & galaxy merger mass ratio between primary (active) and secondary components \\
	32 & Sep{\_}13 & separation between primary (active) and tertiary components [kpc] \\
	33 & Ratio{\_}13 & galaxy merger mass ratio between primary (active) and tertiary components \\
	\enddata	
	\tablecomments{Field numbers 1 -- 29 are taken from the ACS-AGN catalog \citep{Stemo2020}; AGN selection and derivation of AGN host galaxy properties are found therein. Merger selection and derivations of merger properties are described throughout this paper; note that for field numbers 31 and 33 reported values are integrated flux ratios that are used as proxies for merger mass ratio. Lower bound and upper bound are defined as the 16th and 84th percentiles of the distribution, respectively. A ``-999'' value in the table represents no data. The ACS-AGN Merger Catalog is available in its entirety in fits format from the original publisher. \vspace{-12pt}}
	\vspace{-12pt}
\end{deluxetable*}

\subsection{Source Identification and Fitting} \label{subsec: fitting}

Offset and dual AGNs exist in galaxy mergers. In order to find them, we search our active galaxy sample for signs of multiple stellar bulges, which should only exist in galaxy mergers. The dual phase (1 -- 20 kpc stellar bulge separation) of galaxy mergers is especially important because at separations greater than 20 kpc, the galaxies should only be minimally interacting, while at separations less than $\sim$1 kpc, a significant portion of stellar bulges may be coalescing. Therefore, we specifically search for multiple stellar bulges within this range in order to identify offset and dual AGNs.

We can identify which AGN host galaxies contain multiple stellar bulges by modeling them with \texttt{Source Extractor} \citep{Bertin1996} and \texttt{GALFIT} \citep{Peng2002}. Since these galaxies are already known to host at least one AGN, if we identify them as having multiple stellar bulges, we can deduce that these galaxies are undergoing a merger and host an offset or dual AGN. 

In practice, we identified multiple stellar bulges in our active galaxy sample by fitting S\'{e}rsic surface brightness profiles to bright sources in the galaxy tiles. Specifically, we passed the active galaxy tiles through \texttt{Source Extractor}; this identifies bright groupings of pixels in the image as possible sources, fits an elliptical profile to them, and outputs basic information such as the source location, its flux, and its position angle. Given the high redshifts in our sample, \texttt{Source Extractor} is ideal as it detects sources via a threshold method, which is more suited to the detection of low surface brightness objects than peak finding methods \citep{Yee1991}. When detecting sources using \texttt{Source Extractor}, we used a detection threshold of $3\sigma$ above background and a minimum source area of 7 pixels.

We then used the outputs of \texttt{Source Extractor} as initial parameters (source position, source position angle) for fitting with \texttt{GALFIT}. \texttt{GALFIT} minimizes $\chi_{\nu}^{2}$ (reduced $\chi^{2}$) by fitting various modeled surface brightness profiles convolved with a PSF to an input galaxy image. We allowed \texttt{GALFIT} to fit S\'{e}rsic surface brightness profiles to the sources in our tiles; while \texttt{GALFIT} has many more usable surface brightness profiles and can fit highly complex galaxy structure, the limited nature of our method keeps our computational load down while still satisfactorily modeling our science targets (stellar bulges). By varying the S\'{e}rsic index and effective radius, the S\'{e}rsic profile can accurately model an extended bulge structure as well as more compact objects, such as AGNs, when convolved with a PSF. 

The speed at which \texttt{GALFIT} can simultaneously fit multiple profiles to a given image is highly dependent on image size, convolution size, and number of fits. In order to manage the computational load, the sources from \texttt{Source Extractor} were only used as inputs for \texttt{GALFIT} if the source position was within 22 kpc of the center of the image and the source flux was at least 1\% of the maximal source flux within the innermost 10 kpc of the image. The position limitation is reasonable as the AGN host galaxy is approximately centered on our image tile and our goal is to identify offset and dual AGNs separated by 20 kpc or less --- the extra 2 kpc acts as a buffer in case of non-centered tiles. The flux limitation compares possible companions to the central galaxy and is used to reduce the amount of galaxy structure (e.g. spiral arms, star-forming regions, etc.), background galaxies, and other spurious objects fit by \texttt{GALFIT}; this limitation should only restrict selection of the most minor mergers (merger mass ratio $\gtrsim100$). While \texttt{GALFIT} could reproduce nearly all features of a galaxy image given sufficient components and parameters, only stellar bulges are needed to be identified for this work. Lastly, we limited the maximum number of sources being simultaneously modeled to three, as the likelihood of finding systems with four or more stellar bulges is very low and increasing the number of simultaneous fits would result in a large increase in false positives and computation time. Even with all of these restrictions in place, some AGN host galaxies were still fit with over 2600 unique \texttt{GALFIT} models due to the high number of sources present that needed to be fit in all combinations. 

Other restrictions were also put in place to avoid over fitting. Specifically, we restricted our \texttt{GALFIT} components to be centered within 2.5 kpc of its \texttt{Source Extractor} source. This is in addition to default \texttt{GALFIT} restrictions on model parameters that are unphysical (e.g. effective radius $<0.01$ pixel, S\'{e}rsic indices $<$0.01 or $>$20, etc.). This forced \texttt{GALFIT} to model only the sources we wanted to model and to only model those sources with realistic bulge profiles. For our process, this meant that adding spurious S\'{e}rsic components typically resulted in worse fits than modeling with fewer S\'{e}rsic components. Therefore, for each tile we selected the model with the minimum $\chi_{\nu}^{2}$ as the best fit.

Of the 2566 galaxies selected as active for which tiles could be made, 6 could not be modeled due to too many sources being identified by \texttt{Source Extractor} than would be reasonable to computationally model using our methods, with some having as many as 20 sources identified by \texttt{Source Extractor}. These were inspected visually and 2 were selected to be fit using a subset of the sources identified by \texttt{Source Extractor} that appeared to be possible stellar bulges. In all, 350 active galaxies were best fit by a \texttt{GALFIT} model with two or three components, while the rest were either best fit with a single S\'{e}rsic component or with no S\'{e}rsic component (i.e. only a background component); the ones best fit with no S\'{e}rsic component typically had extremely dim surface brightness profiles.

\subsection{False Positive Reduction} \label{subsec: false pos}

Not all 350 galaxies that were best fit with two or three components by \texttt{GALFIT} truly contain multiple stellar bulges. Even with the initial source limitations imposed prior to  \texttt{GALFIT} modeling, there were still a significant number of models with components fit to non-bulge features (e.g. spiral arms, star-forming regions, etc.). In order to reduce these false positives, we put in place further restrictions on our \texttt{GALFIT} component parameters: a centroid separation limit, an integrated flux ratio limit, and a significance above background restriction.

When \texttt{GALFIT} models an image, there is no inherent lower limit to component separation and it is only restricted in its upper limit to the size of the image. These limits are unrealistic on the low end and not useful for this work on the high end. Since our goal is to identify offset and dual AGNs, the separation of possible stellar bulges containing SMBHs should be inherently limited to $<$20 kpc.  However, there must also be a lower limit; an absolute lower limit would be the angular resolution of the observing instrument (0\farcs05 for \textit{HST}/ACS), but the lower limit of typical bulge sizes is actually the more restrictive lower separation in this case. For example, the nearby ($z = 0.02$) active dwarf galaxy RGG 118 has a bulge diameter of approximately 1.5 kpc \citep[][]{Baldassare2017}. This 1.5 kpc lower limit is a conservative estimate for possible bulge sizes identifiable in our sample; the moderate to massive galaxies of the ACS-AGN should contain similar or larger bulges, and 1.5 kpc is above the resolution limit of \textit{HST} for all redshifts present in our sample. Therefore we impose a separation limit for our modeled stellar bulges of $>$1.5 kpc and $<$20 kpc.

In addition, the automated nature of our bulge fitting method frequently results in additional components being fit to non-bulge features such as spiral arms, star-forming regions, and occasionally low-surface brightness, extended features. In order to address these issues, we put in place two further restrictions on the modeled \texttt{GALFIT} components. First, we required that the integrated fluxes of any paired components must be within 5 magnitudes of each other (1\%). If we take the integrated flux ratio as a proxy for the stellar mass ratio (see Section \ref{subsec: merge props}), this constraint limits the possible merger mass ratios of our candidates to $<$100. This restriction is successful at eliminating many false positives typically related to star-forming regions and some spiral arms and only eliminates the most minor of mergers. Our second restriction is a significance above background criterion; specifically, we require that the modeled flux at the centroid of any fit component must be at least 5$\sigma$ above the background, as modeled by \texttt{GALFIT}. This restriction is excellent at eliminating more extended, low-surface brightness false positive fits, such as those typically associated with diffuse gas features and some spiral arms.

After we removed a large number of false positives through automated means, 280 active galaxies were still identified as hosting multiple stellar bulges. Due to the automated nature of our approach, there were still some false positives included in this set. Therefore, as a final check, this reduced sample was visually examined independently by the three lead authors of this work, with any split determinations resulting in the conservative decision of removal from the sample. We also used this opportunity to refit individual stellar bulge models as needed in order to more accurately model the offset and dual AGN system; this was typically slightly adjusting the centroid of the bulge fit, the axis ratio,  or the S\'{e}rsic index and effective radius to better reflect the observed data. This final verification and refit resulted in 220 galaxies identified as offset and dual AGNs --- 199 of these being two-bulge systems and 21 being three-bulge systems. Some examples of offset and dual AGNs that we identified are shown in Figure \ref{plot: examples}.

For the rest of this paper, we will analyze the two-bulge systems and the galaxy pairs containing the AGN host galaxy and the brightest secondary galaxy in the three-bulge systems. While this choice of analyzing only the brightest pairs in triple systems could theoretically lead to a bias in our analysis, we verified that none of the analyses or results presented in the remainder of the paper would be significantly altered if we were to examine all pairs (i.e. both sets of pairs containing the AGN host galaxy in triple systems) or if we were to completely exclude triple systems from any analysis. Differentiating offset from dual AGNs in our sample and analysis of the individual sub-populations, as well as the triple bulge systems, will be the focus of a future paper (Stemo et al., in prep).

\section{Bias Analysis \& Correction} \label{sec: bias}
All methods of identifying galaxy mergers are prone to significant selection biases. The process outlined in Section \ref{sec:Bulge Select} relies upon the ability to properly detect and separate the stellar bulges associated with the AGNs or SMBHs of each merger component, which can be extremely challenging. 

As the bulges near each other, their surface brightness profiles can overlap significantly; but as was discussed in Section \ref{subsec: false pos}, bulges can be as close as 1.5 kpc before they physically begin to coalesce. This means that physically distinct stellar bulge pairs at small separations can be especially difficult to detect even with \textit{HST}/ACS 's spatial resolution. This is exacerbated by the high redshift nature of our sample, with a 1.5 kpc separation corresponding to less than 6 pixels in the most extreme region of our redshift range.

In addition to the challenge in resolving small separations, the higher the merger mass ratio (the more dissimilar the masses), the more likely the system is to be modeled as a single component rather than as multiple stellar bulges. As the merger mass ratio increases, the integrated flux ratio also increases. This leads to one surface brightness profile dominating the other; therefore, the model of the lower integrated flux (lower mass) component is less important to the goodness of fit of the overall model.

Both of these factors cause our merger identification pipeline to frequently incorrectly model closely separated and high merger mass ratio galaxy mergers with a single component. This means that our galaxy merger sample is biased towards major mergers (low mass ratio) and mergers at large separations. In order to correct for this, we simulated and modeled 377,000 galaxy mergers across a large swath of parameter space. Examining the resulting \texttt{GALFIT} models of these simulated mergers allowed us to correct for the aforementioned separation and merger mass ratio selection biases.

In Section \ref{subsec: create sims}, we discuss the creation of the galaxy merger models and the parameter space they cover. In Section \ref{subsec: model sims}, we detail the \texttt{GALFIT} modeling of the simulated mergers and report the regions of success and failure of our merger identification pipeline. Lastly, in Section \ref{subsec: bias correction}, we discuss the quantification of our selection biases and the process we used to correct for them.

\begin{figure}[!t]
	\begin{center}
		\includegraphics[width=0.45\textwidth]{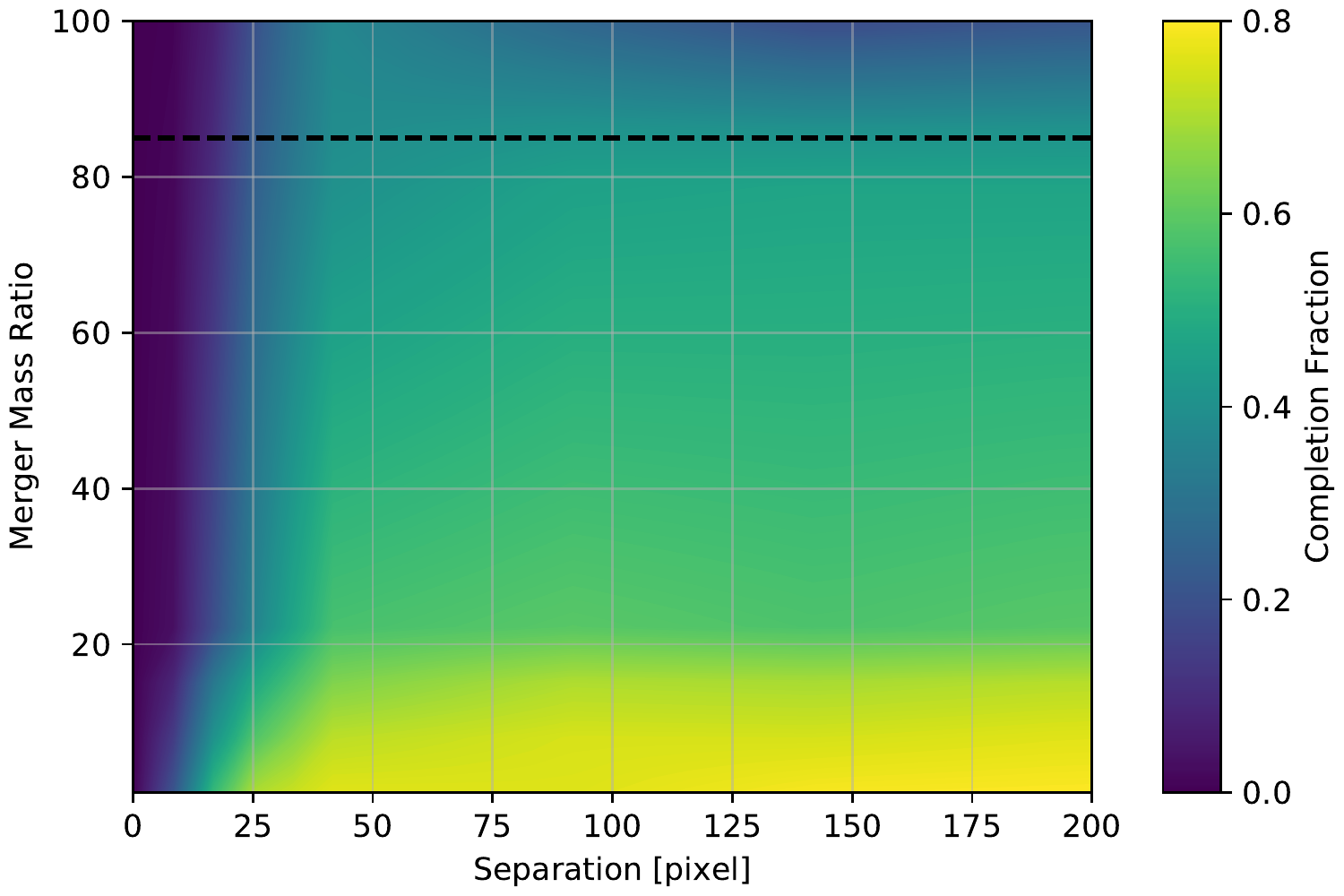}
	\end{center}
	\vspace{-12pt}
	\caption{The bias map of our merger identification pipeline as determined by simulating and fitting S\'{e}rsic pairs across a large span of parameter space. The black dashed line is the upper merger mass ratio limit ($=85$) where this bias map is accurate. Note how quickly and severely the selection becomes biased at small separations.}\label{plot: bias map}
	\vspace*{-8pt}
\end{figure}

\subsection{Simulating Galaxy Mergers}\label{subsec: create sims}

Our merger identification pipeline relies on modeling multiple stellar bulges with S\'{e}rsic profiles. Ignoring position, S\'{e}rsic profiles only need three variables to be fully defined; these are typically the S\'{e}rsic index, effective radius, and intensity at the effective radius. However, we choose to define integrated flux (i.e. total flux) as the last variable instead of intensity at the effective radius for our simulations. Since we are modeling mergers, we also need to define a merger mass ratio and a separation. 

In order to properly estimate the biases involved with our approach to identifying galaxy mergers, we created a large number of simulated S\'{e}rsic pairs with parameter values that span the expected parameter space, as estimated by our observed \textit{HST} galaxies, convolved them with an appropriate PSF for the ACS instrument on \textit{HST}, and then added noise to the data which mimicked the observed background of our galaxy tiles. 

We simulated all combinations of the following parameter arrays: S\'{e}rsic index = [0.25, 0.5, 1, 3, 6, 10], effective radius = [2, 5, 10, 15, 30, 60, 100], integrated flux = [$10^{3}$, $10^{4}$, $10^{5}$, $10^{6}$, $10^{7}$, $10^{8}$], merger flux ratio = [1, 2, 4, 8, 16, 40, 100], and separation = [3.5, 10, 15, 25, 50, 100, 150, 196]. Here we are using the merger flux ratio as a proxy for the merger mass ratio. The effective radius and separation are in units of pixels (\textit{HST}/ACS has a pixel scale of 0\farcs03 per pixel), while the integrated flux is in units of ADUs. The separations were chosen to span the range of 1 to 20 kpc across our entire sample redshift range, while the integrated fluxes were chosen to fully span the range of our observed sample integrated fluxes. We restricted our simulated mergers by limiting the ratio of effective radius to S\'{e}rsic index to being less than 40; this was done because galaxy parameters outside of that are not physical. To best mimic the observed background of our galaxy tiles, we added Poisson noise from our simulated models as well as Gaussian noise with a mean of 1200 and standard deviation of 400; these values were estimated from the distribution of our sample's \texttt{GALFIT} modeled backgrounds.

 In total we created 376,992 simulated galaxies to model, consisting of 11,088 simulated single galaxies and 365,904 simulated galaxy pairs.

\begin{figure*}[t]
	\centering
	\includegraphics[width=0.47\textwidth]{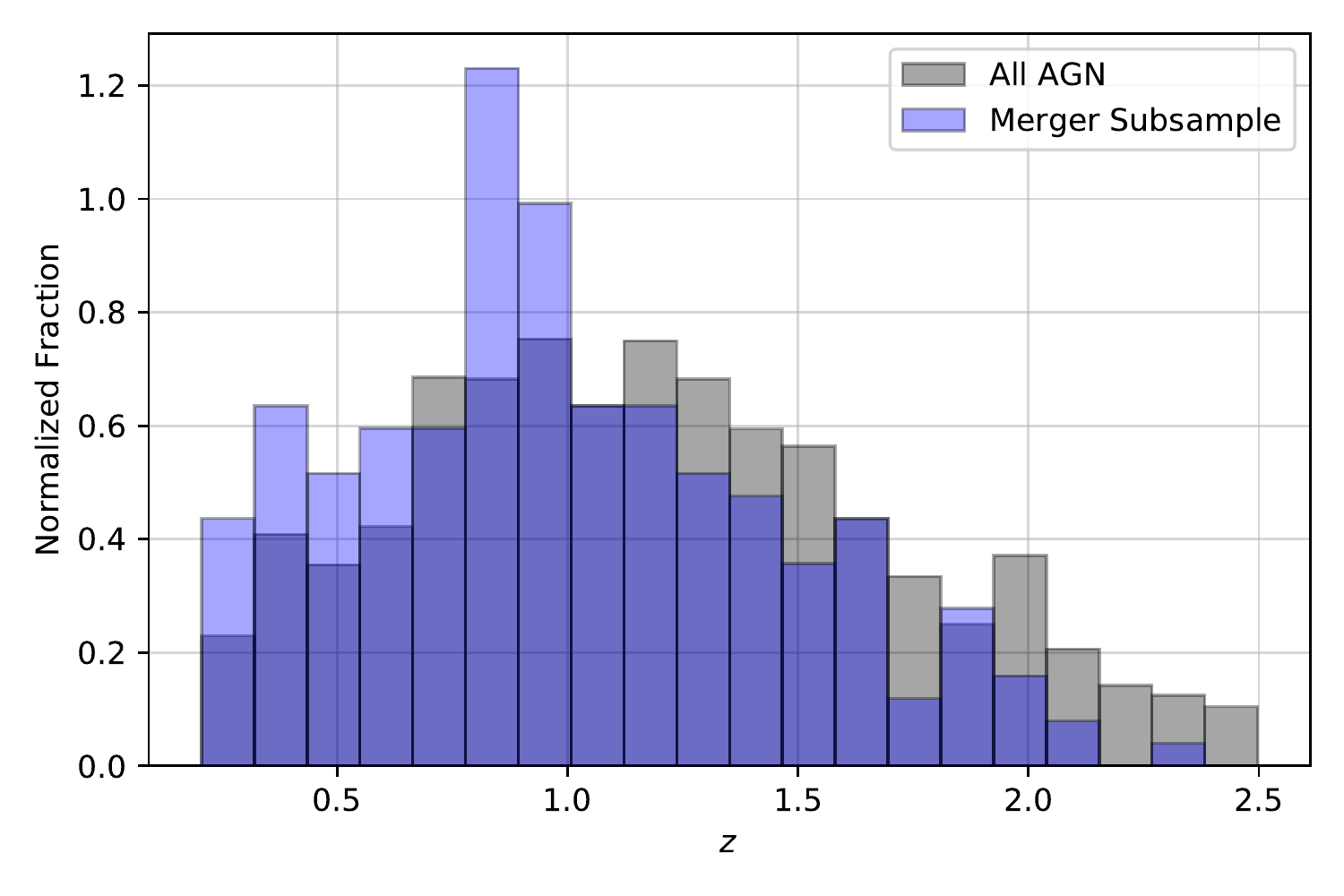}
	\vspace{-6pt}
	\includegraphics[width=0.47\textwidth]{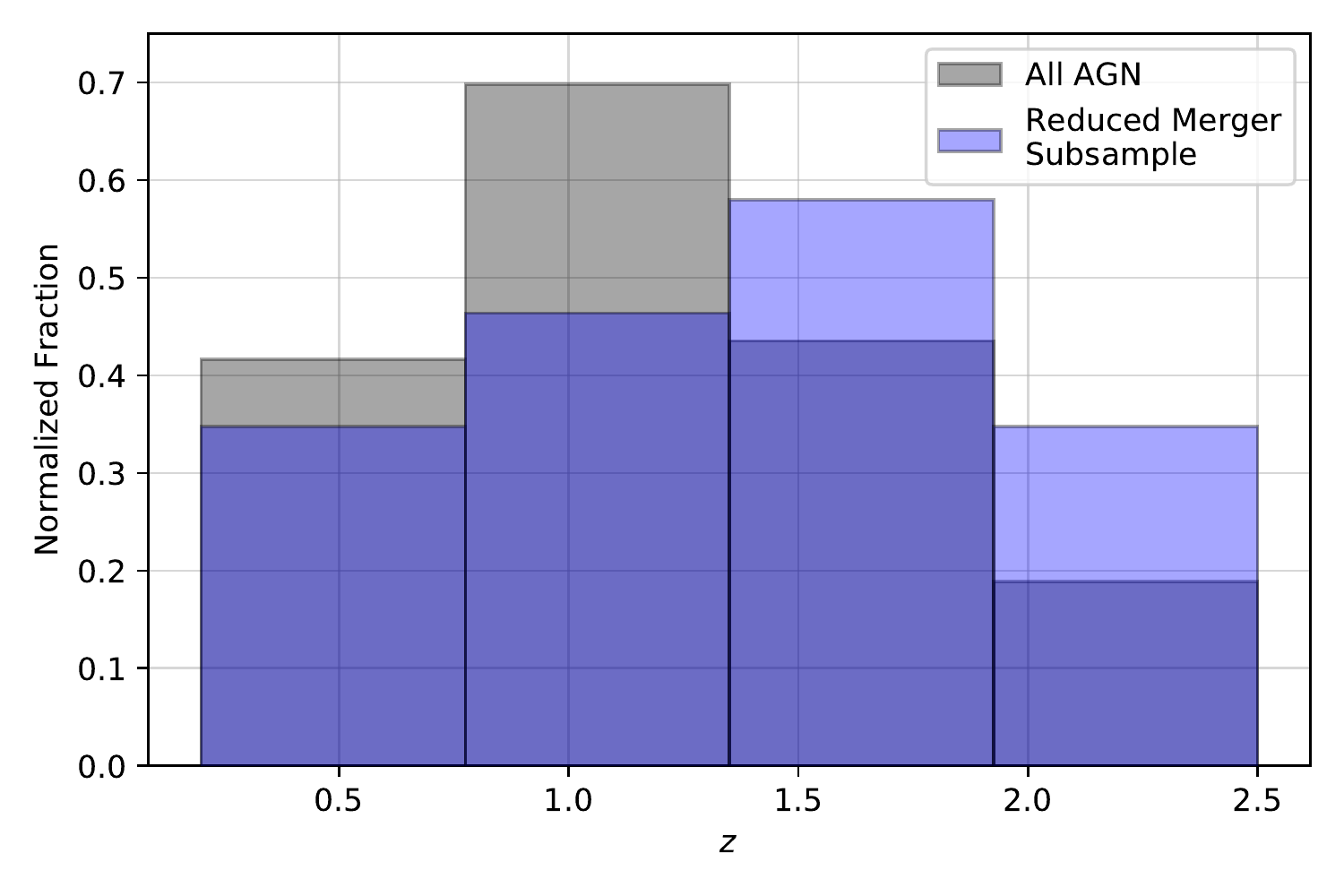}
	\includegraphics[width=0.47\textwidth]{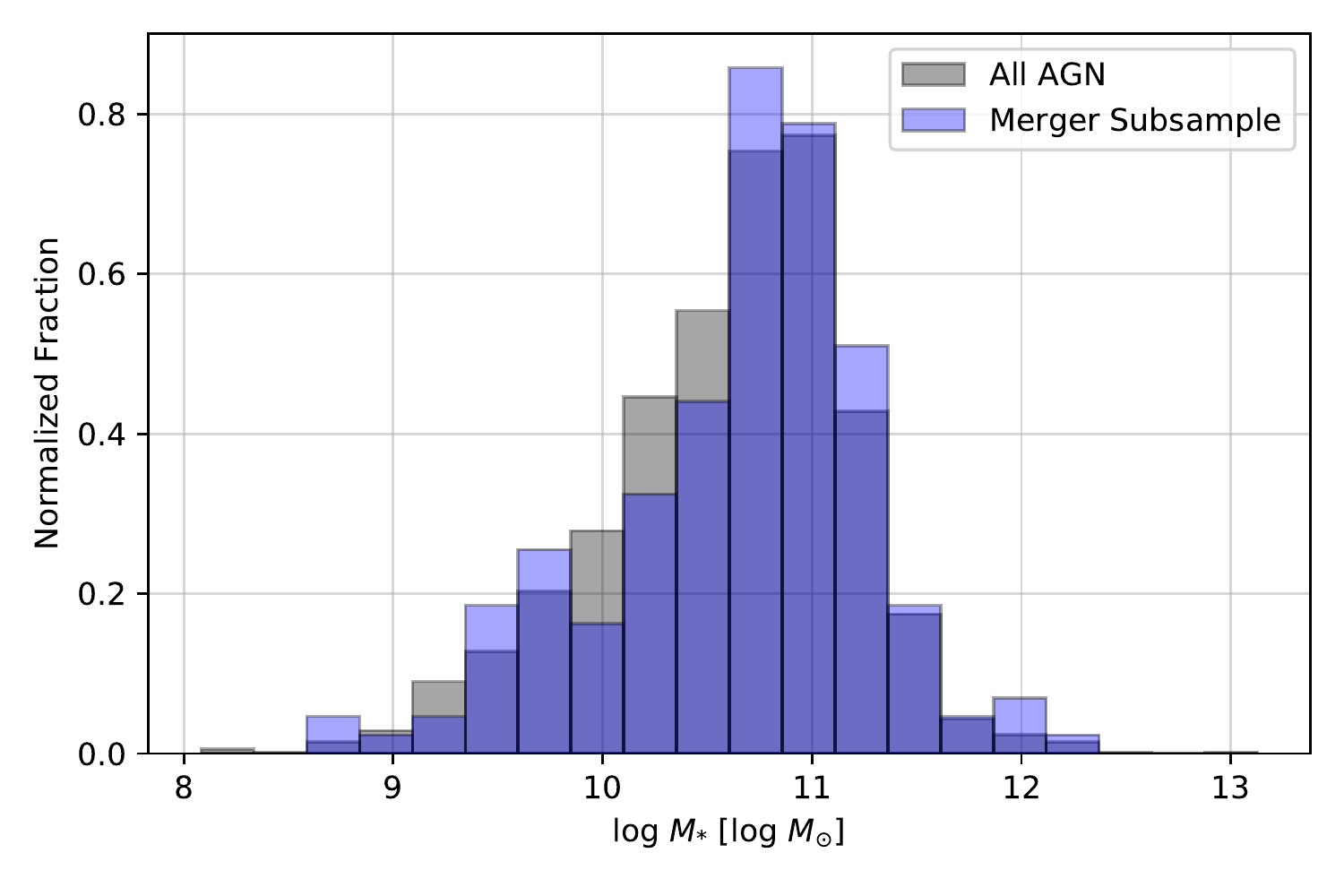}
	\vspace{-6pt}	
	\caption{Histogram of the redshifts (top left) of our merger subsample in comparison to the general AGN population of the ACS-AGN Catalog; note that the histograms are normalized to an area of 1 for comparison purposes. The over-representation of offset and dual AGNs at low redshift and under-representation at high redshift is due to selection effects. Once these effects have been accounted for, the expected distribution of mergers as a function of redshift is found, as seen in the histogram of the bias-corrected merger subsample (top right, Section \ref{subsec: acs-agn compare}). Below these we see the histogram of the stellar masses (bottom) of our merger subsample in comparison to the general AGN population of the ACS-AGN Catalog: the merger subsample distribution does not differ significantly from that of the general AGN population.}
	\label{plot: z plots}
	\vspace{6pt}
\end{figure*}

\subsection{Modeling the Simulated Mergers}\label{subsec: model sims}

We then ran these simulated galaxies through our merger identification pipeline as detailed in Sections \ref{subsec: fitting} \& \ref{subsec: false pos}. Of the 11,088 simulated single galaxies, 5565 were correctly modeled, 5443 were not modeled at all, and 80 were incorrectly modeled with only a background component. The 5443 that were not modeled had low intensity levels; specifically they had intensities (in units of ADUs) at their effective radius of $I_{e}<10^{0}$ and central intensities of $I_{0}<10^{2.7}$. The 80 that were incorrectly modeled had $I_{e}>10^{0}$ and $I_{0}>10^{2.7}$, and therefore can be classified as false negatives. Of the 5565 true positives, all had $I_{e}>10^{0}$ and $I_{0}>10^{2.7}$.

Of the 365,904 simulated galaxy pairs, 64,979 were correctly modeled, 180,724 were not modeled at all, and 120,188 were incorrectly modeled with either a single galaxy component or only a background component. There were only 13 simulated pairs that were modeled with more than two galaxy components. The 180,724 that were not modeled had $I_{e}<10^{0}$ and $I_{0}<10^{2.7}$, while 69,526 of the 120,188 that were incorrectly modeled had $I_{e}>10^{0}$ and $I_{0}>10^{2.7}$, and therefore can be classified as false negatives. Of the 64,979 true positives, all had $I_{e}>10^{0}$ and $I_{0}>10^{2.7}$.

\subsection{Correcting for Selection Biases}\label{subsec: bias correction}

In order to properly account for the biases in our selection pipeline, we calculated the fraction of true positives to true positives plus false negatives; there were 70,544 true positives out of 140,150 true positives and false negatives; we define this as the completion fraction. We calculated the completion fraction at all points across the separation and merger mass ratio axes, which gave us a sparse two-dimensional bias map that we interpolated (Figure \ref{plot: bias map}) in order to allow us to correct our results across the entire separation and merger mass ratio range of our sample. The simulations with $I_{e}<10^{0}$ and $I_{0}<10^{2.7}$ were considered true negatives (i.e. accurately not found due to being too faint for the background noise --- this mirrors a sensitivity limit for true data), while the 13 false positives were ignored as their contribution is trivial to the overall statistics (i.e. less than a 0.001 contribution to the completion fraction).

While examining our biases across our parameter space, we found that when the separation exceeded 50 pixels at the highest merger mass ratio (1:100), our pipeline's accuracy dropped significantly. This is due to the \texttt{Source Extractor} source flux limitation we used. Since \texttt{Source Extractor} limits the flux ratio between multiple sources to 1:100 prior to modeling by \texttt{GALFIT}, our pipeline becomes insensitive to merger mass ratios $\gtrsim85$. This does not affect our analysis since we have no offset or dual AGNs with merger mass ratios greater than 85, but we include this discussion for completeness.

We applied corrections for these biases to our sample. When examining how AGN activation behaves as a function of bulge separation (see Section \ref{subsec: result AGN frac}), we first binned by redshift in order to assign pixel separations to our kpc separation bins, then binned into our kpc bins, and lastly binned by merger mass ratio. We then applied a correction factor, the reciprocal of the completion factor calculated at the mean separation and merger mass ratio values of our bins, to offset our biases; we also estimated error values by calculating this correction at the minimum and maximum separation and mass ratio values in each bin. 

We repeated this process to examine how AGN activation relates to merger mass ratio (see Section \ref{subsec: result AGN major}); the only difference was the binning order, with merger mass ratio first, followed by bulge separation. Lastly, if the error in a bin was less than the Poisson statistic of $\sqrt{N}$, where $N$ was the number added as a correction in that bin, we set the error to $\sqrt{N}$; this was done to account for any errors associated with our assumptions as well as unknown errors so as to not overestimate the accuracy of our methods.

\section{Merging Galaxy Properties \& \\ Comparison to ACS-AGN Sample} \label{sec: gal props}

The ACS-AGN Catalog contains AGN and host galaxy properties, most derived from SED template fits to photometric data; these data include: redshift ($z$), AGN bolometric luminosity ($L_{AGN}$), galaxy mass ($M_{*}$), star formation rate (SFR), and column density ($N_{H}$). The process by which these properties are derived from observations or from SED template fits is explored in depth in S20. The selection and modeling of our offset and dual AGN sample in this work allows us to derive some merger specific properties, such as bulge separation and merger mass ratio.

In Section \ref{subsec: merge props}, we discuss how we calculate these merger specific properties. We then examine the AGN and host galaxy properties of offset and dual AGNs in comparison to the general AGN sample of the ACS-AGN Catalog in Section \ref{subsec: acs-agn compare}. Lastly, we search for any correlations between the AGN, host galaxy, and merger properties of offset and dual AGNs in Section \ref{subsec: props vs merge param}.

All of the properties discussed in this section, and more, are included in the ACS-AGN Merger Catalog, presented in Table \ref{tab: catalog}. This catalog is also available in its entirety in fits format from the original publisher.

\begin{figure*}[!]
	\centering
	\includegraphics[width=0.47\textwidth]{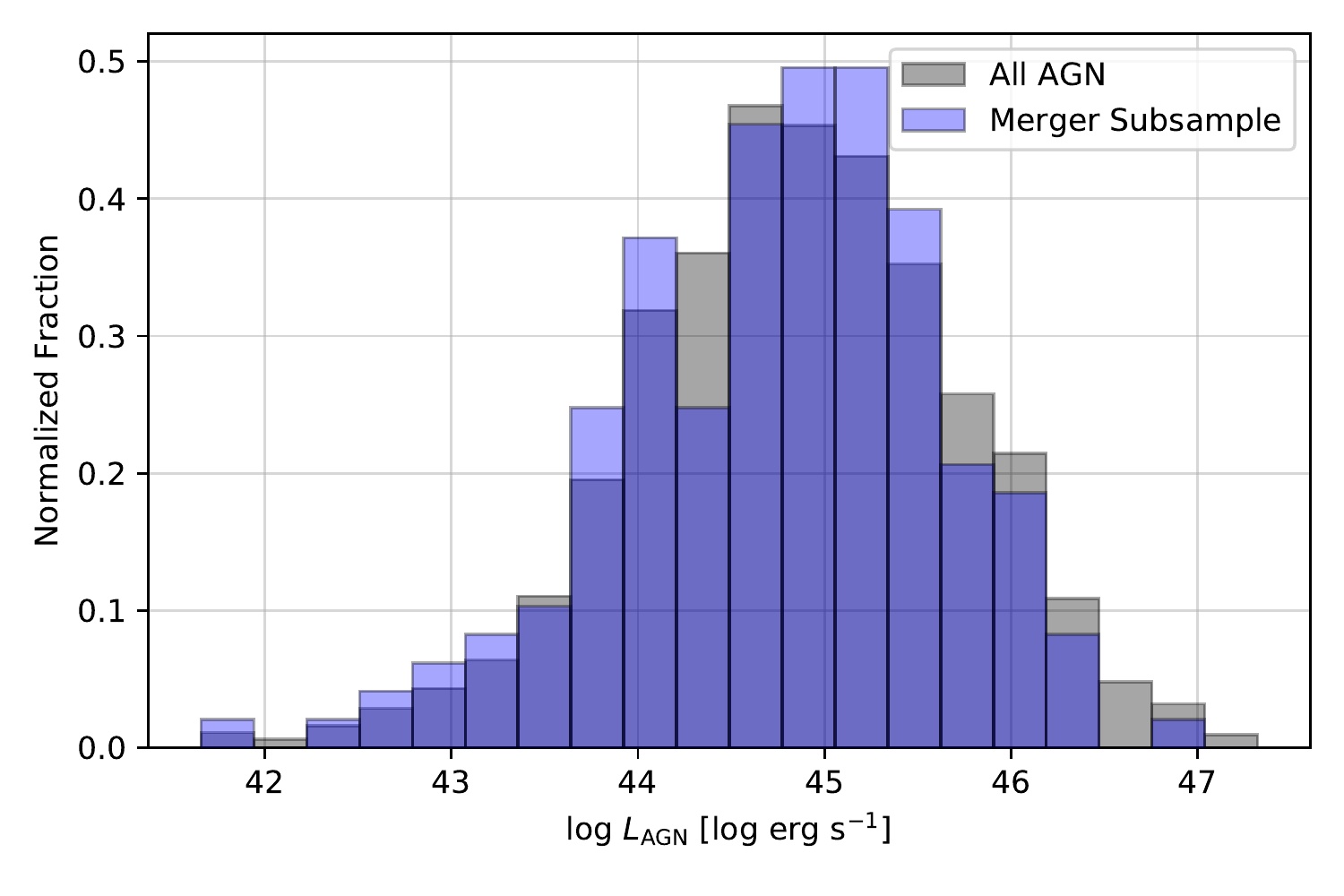}
	\vspace{-6pt}
	\includegraphics[width=0.47\textwidth]{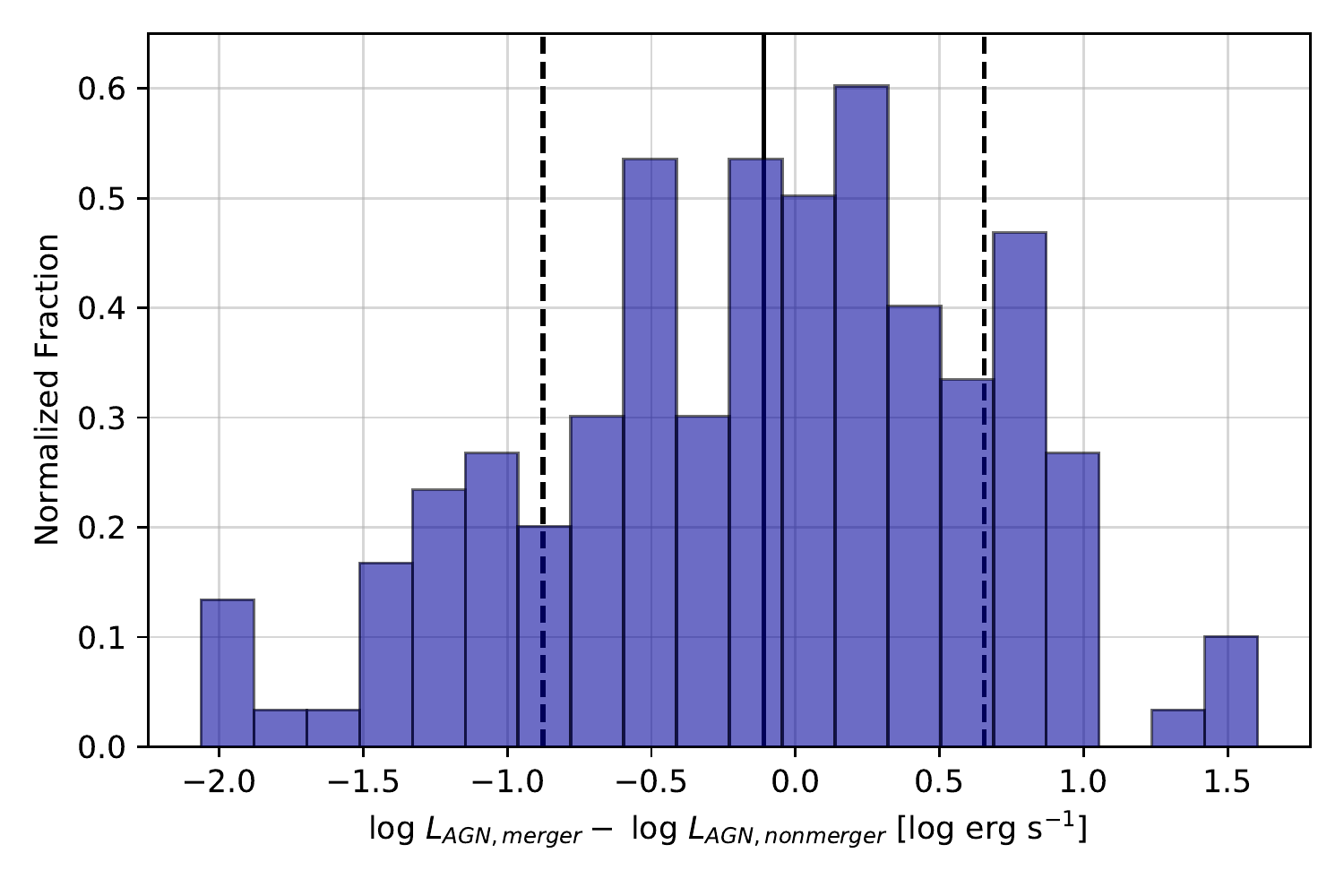}
	\includegraphics[width=0.47\textwidth]{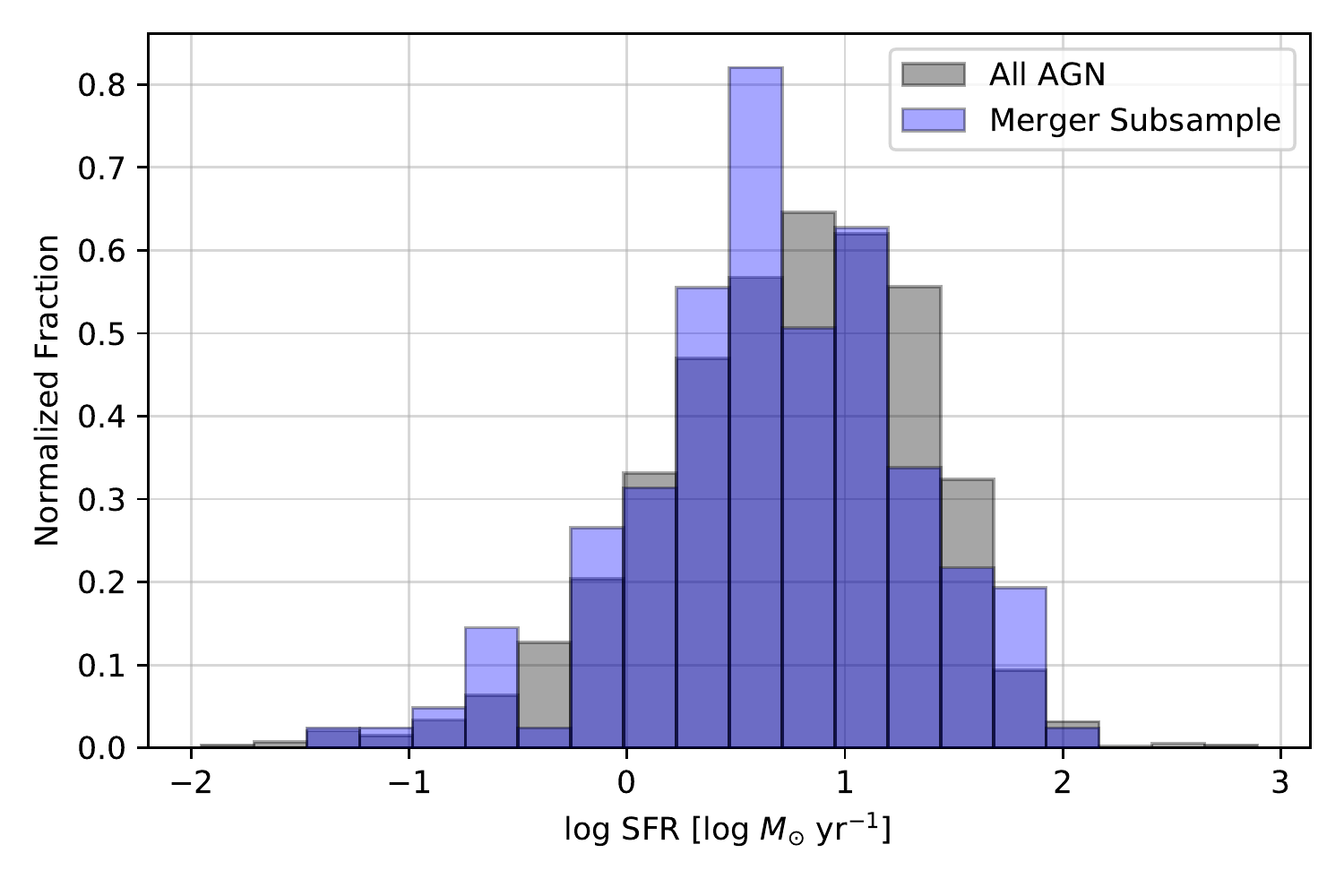}
	\vspace{-6pt}
	\includegraphics[width=0.47\textwidth]{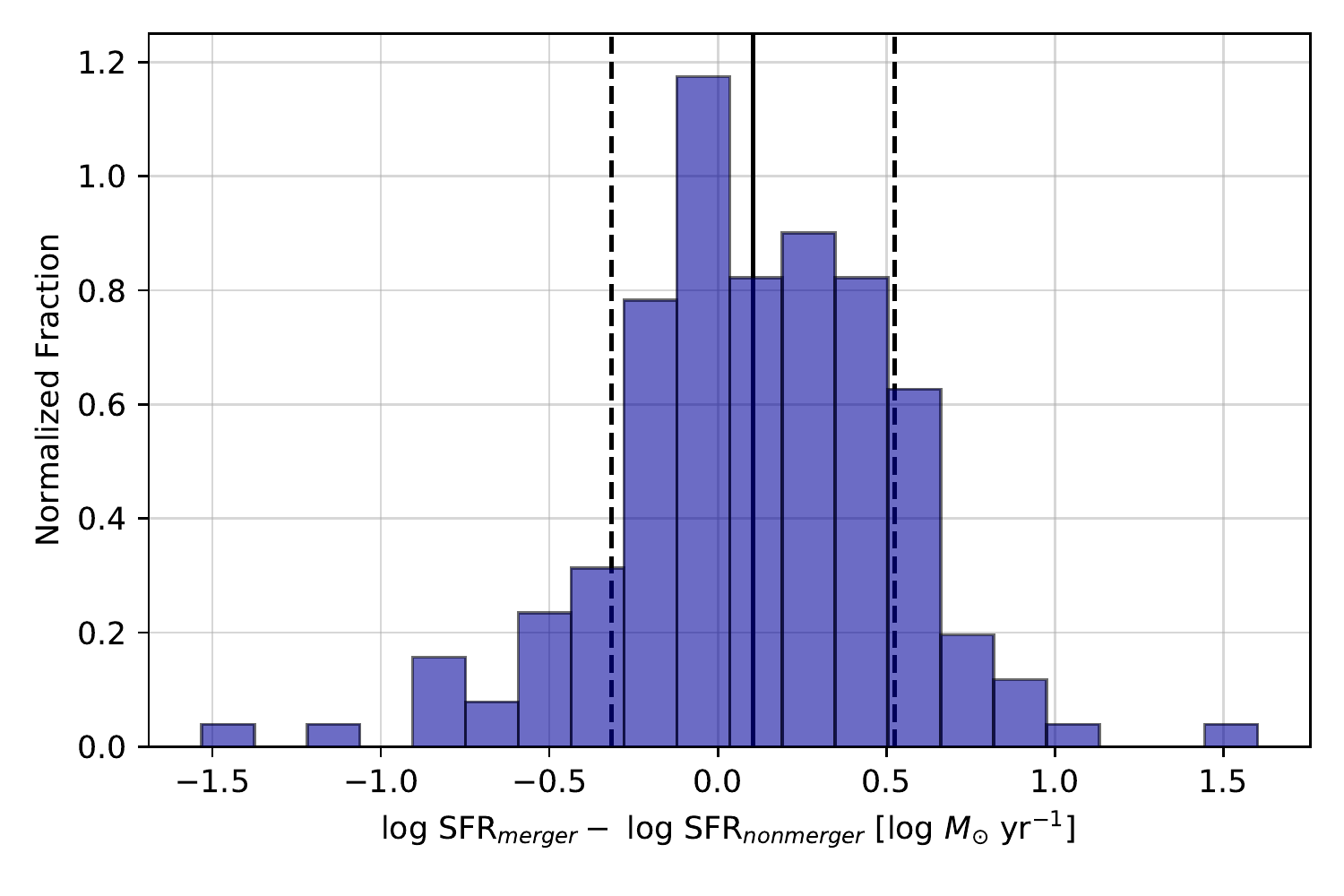}
	\includegraphics[width=0.47\textwidth]{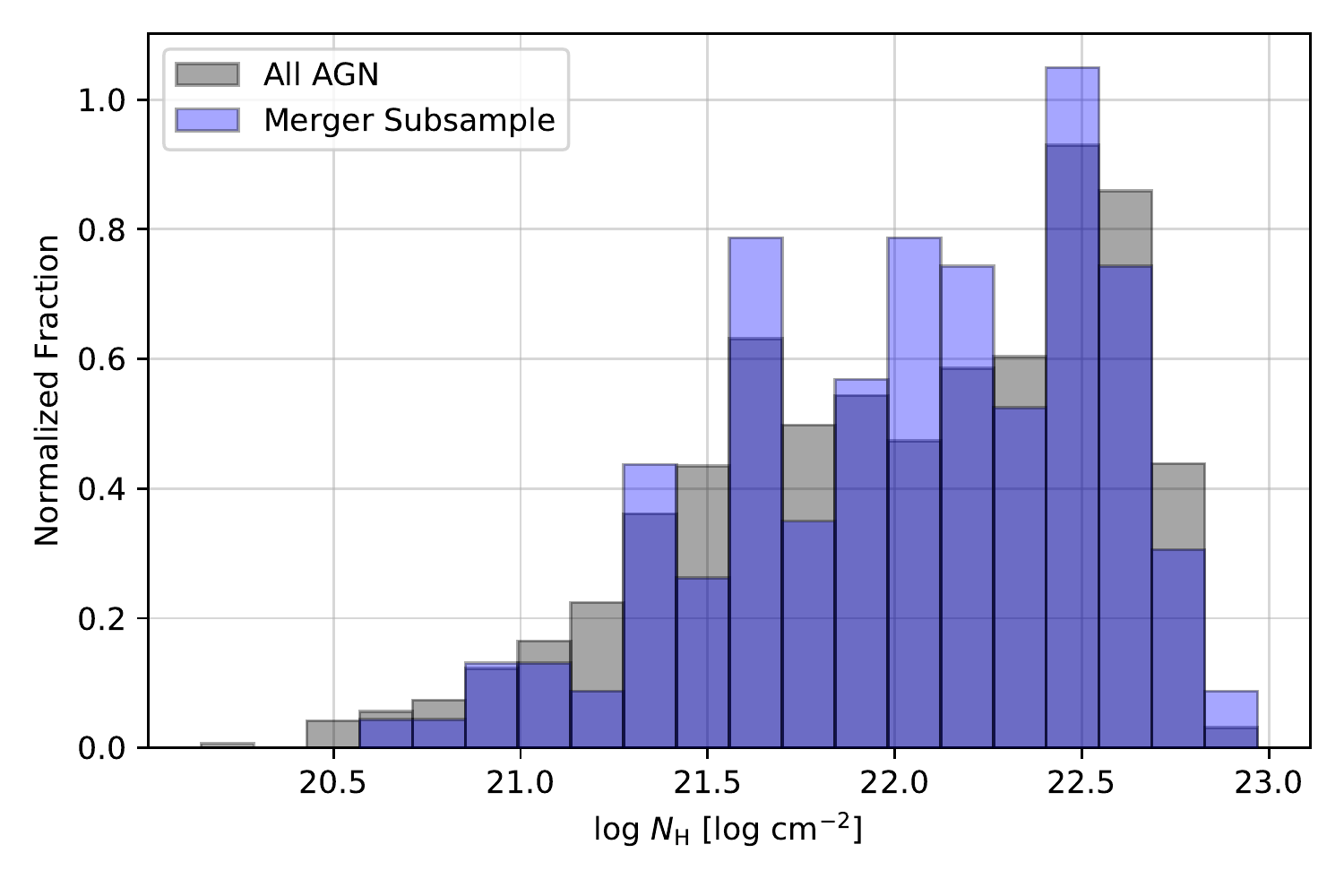}
	\includegraphics[width=0.47\textwidth]{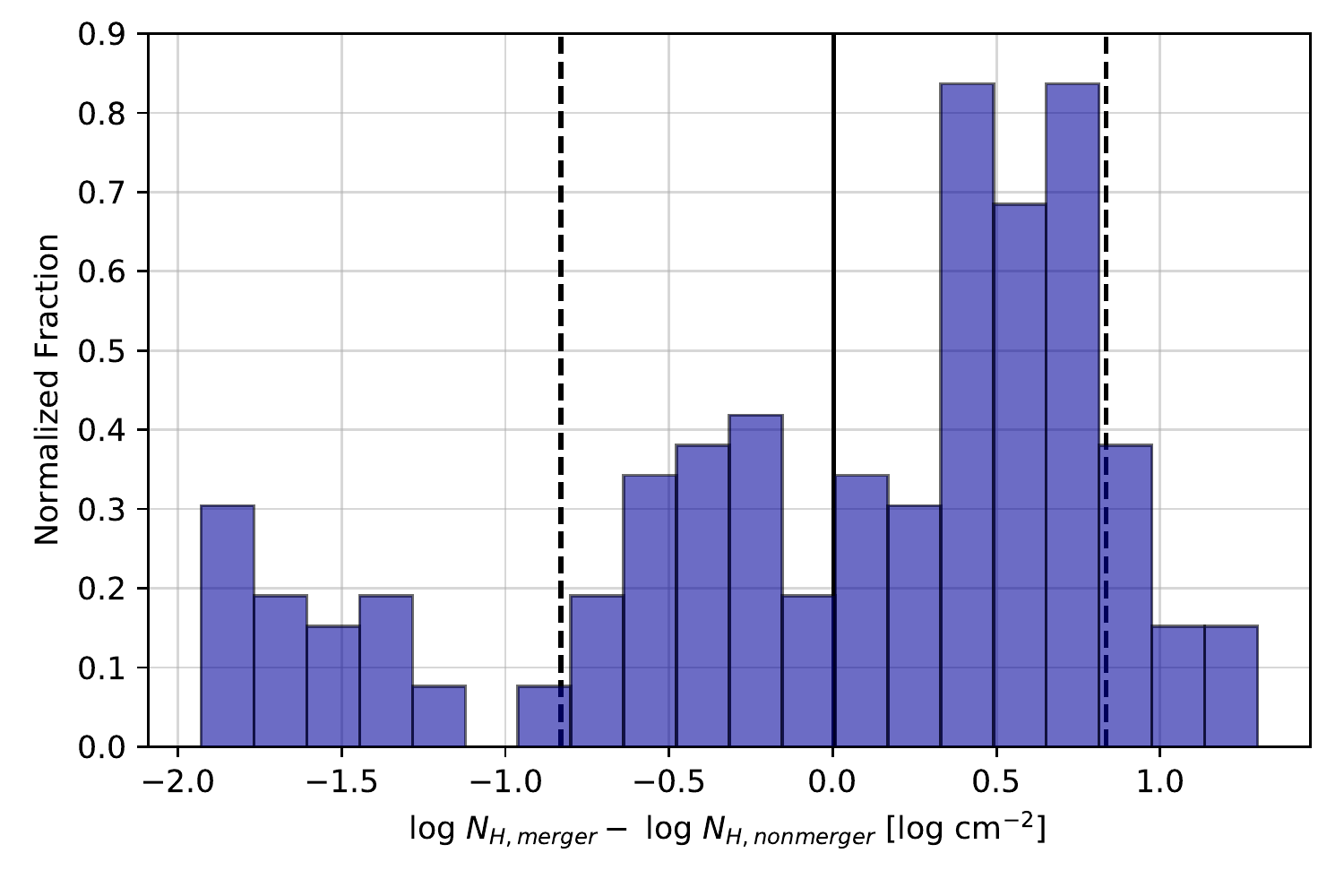}	
	\vspace{-6pt}		
	\caption{Histograms of the AGN and host galaxy property values of our merger subsample in comparison to the general AGN population of the ACS-AGN Catalog (left) alongside histograms of the difference in AGN and host galaxy properties of our merger subsample in comparison to non-merging AGN host galaxies matched in redshift and galaxy stellar mass (right); the vertical solid lines indicate the mean difference values while the vertical dashed lines indicate the difference distributions' standard deviations. Note that the merger subsample distributions mimic that of the general AGN population and that the means of the difference distributions are approximately zero.}
	\label{plot: gal props 1}
	\vspace{6pt}
\end{figure*}

\subsection{Obtaining Merger Parameters}  \label{subsec: merge props}

Two important merger parameters can be obtained from the \texttt{GALFIT} models: component separation and the integrated flux of each component. The centroid locations of all fit components are given by \texttt{GALFIT} in image units (pixel values). Using these centroid locations, and knowledge of the pixel scale of the galaxy tiles, we back out the angular separation of the stellar bulges for our candidates. The angular separation is then transformed into a projected physical separation using the angular distance calculated from each galaxy's redshift.

The second significant merger parameter that can be obtained from the \texttt{GALFIT} model is merger mass ratio. \texttt{GALFIT} measures and reports an integrated flux measurement for each component during its fitting process. We take the ratio of integrated fluxes between bulge pairs as a proxy for their stellar mass ratio. 

The distributions of these parameters are shown in Figures \ref{plot: sep histogram} \& \ref{plot: ratio histogram}, respectively, and will be discussed further in Section \ref{sec: results}.

\begin{figure*}[!]
	\centering	
	\includegraphics[width=0.48\textwidth]{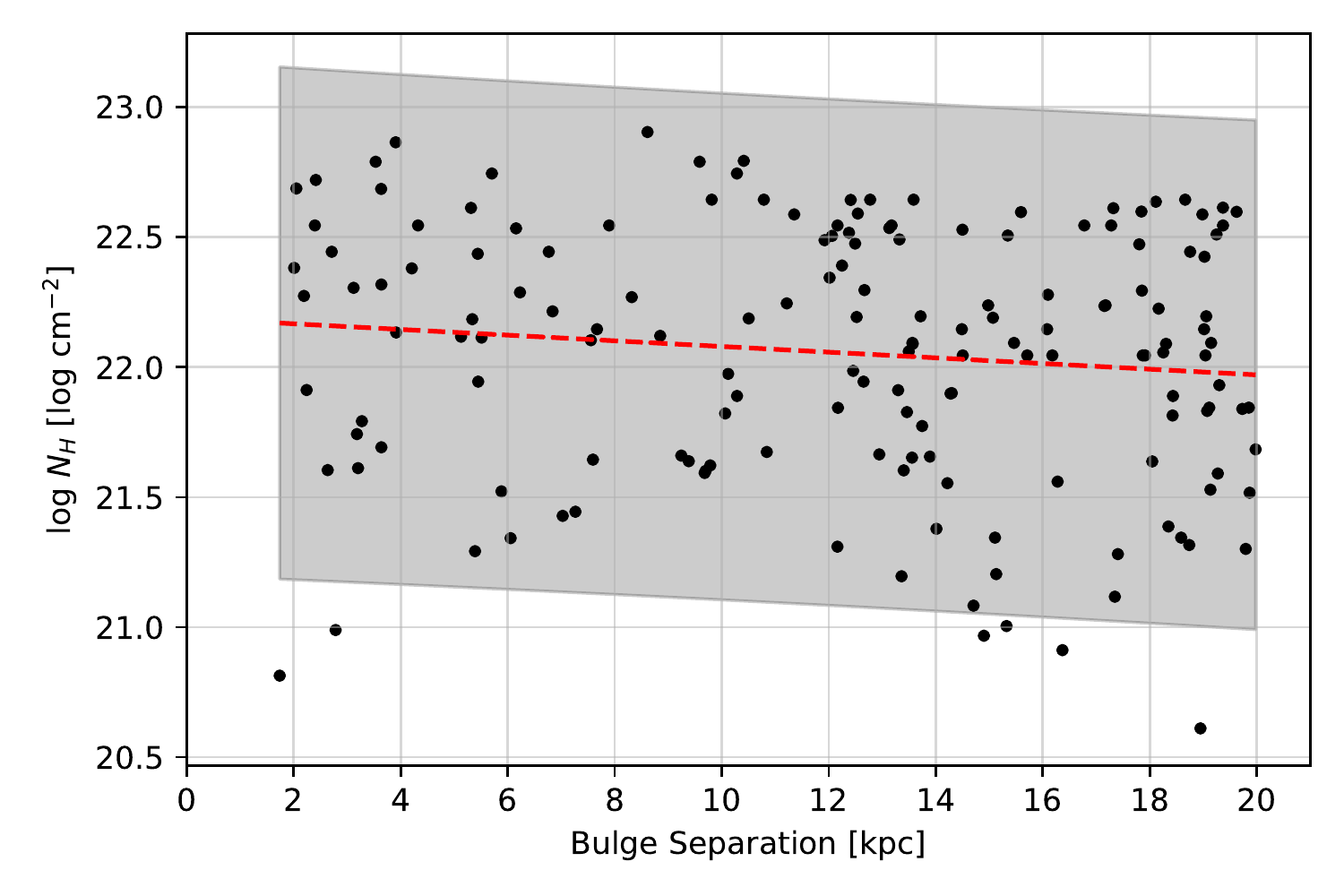}
	\vspace{-6pt}
	\includegraphics[width=0.48\textwidth]{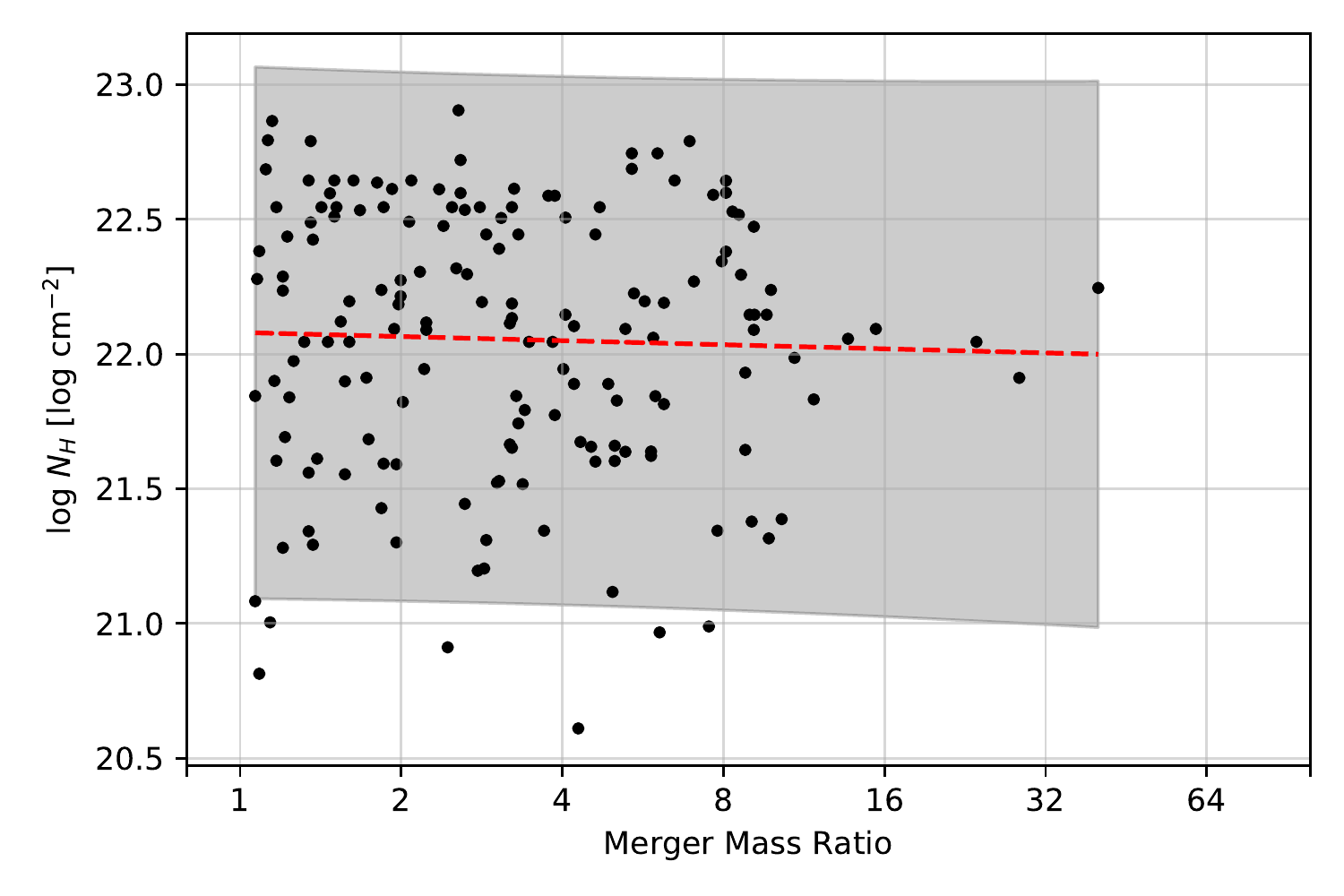}
	\includegraphics[width=0.48\textwidth]{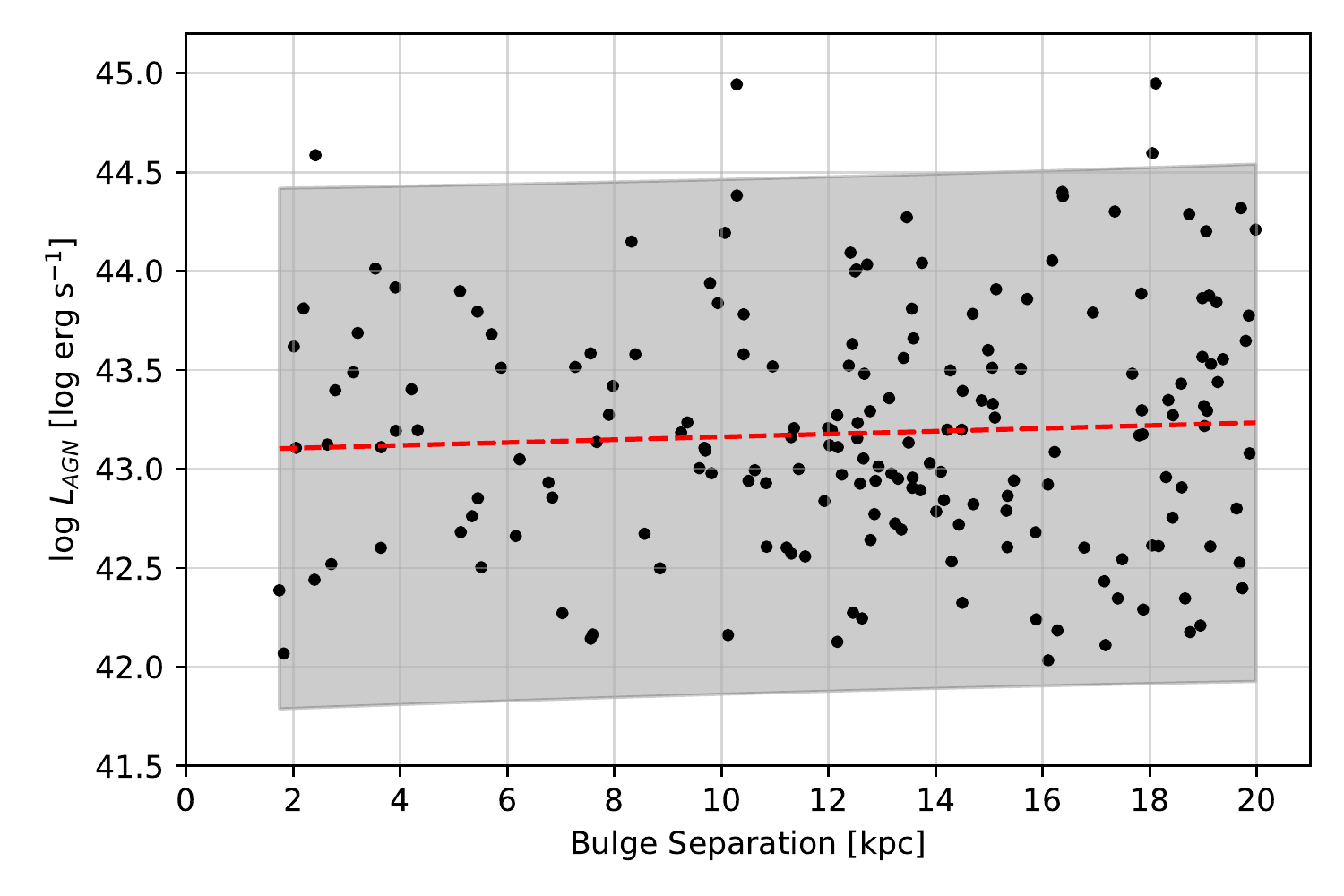}
	\vspace{-6pt}
	\includegraphics[width=0.48\textwidth]{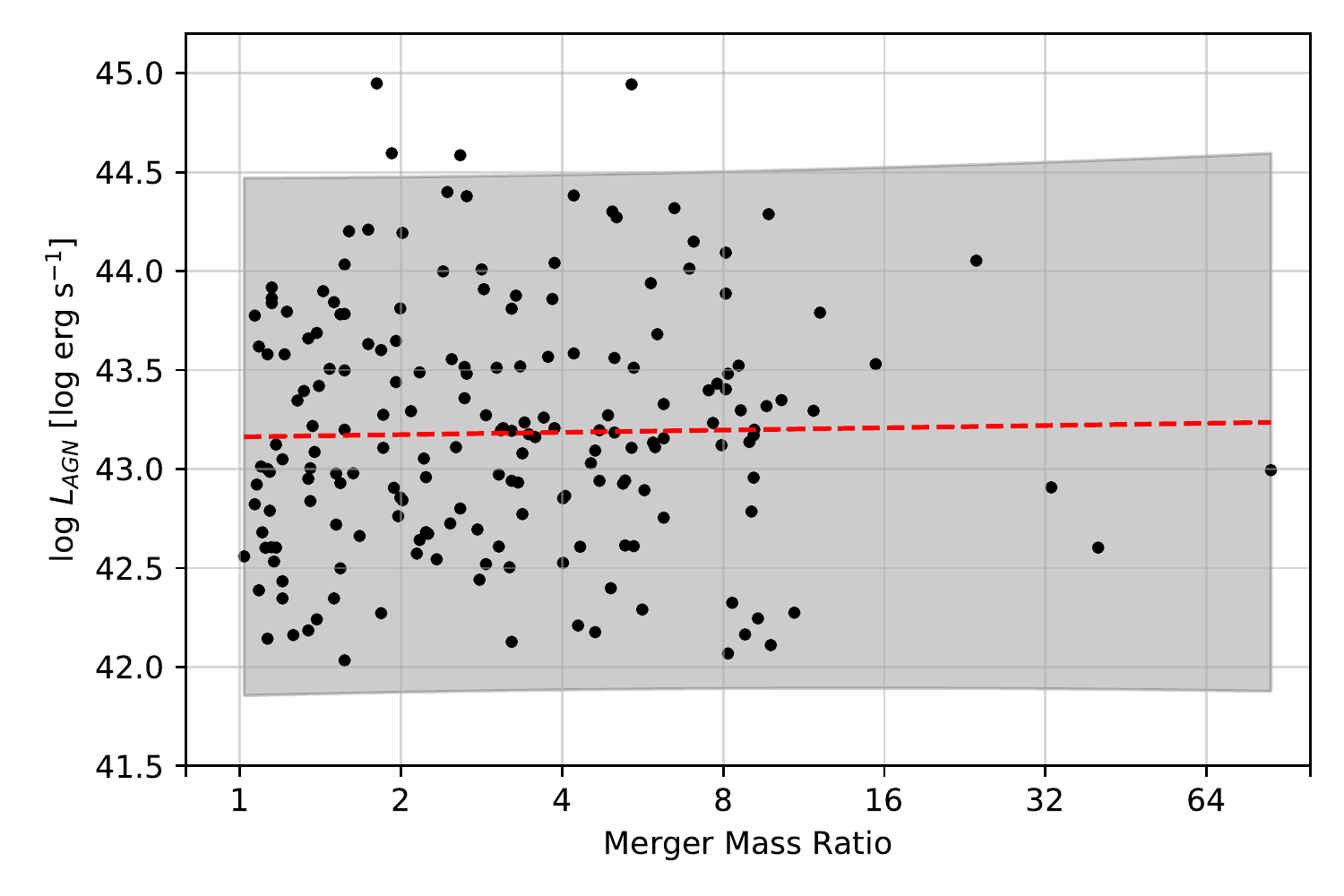}
	\includegraphics[width=0.48\textwidth]{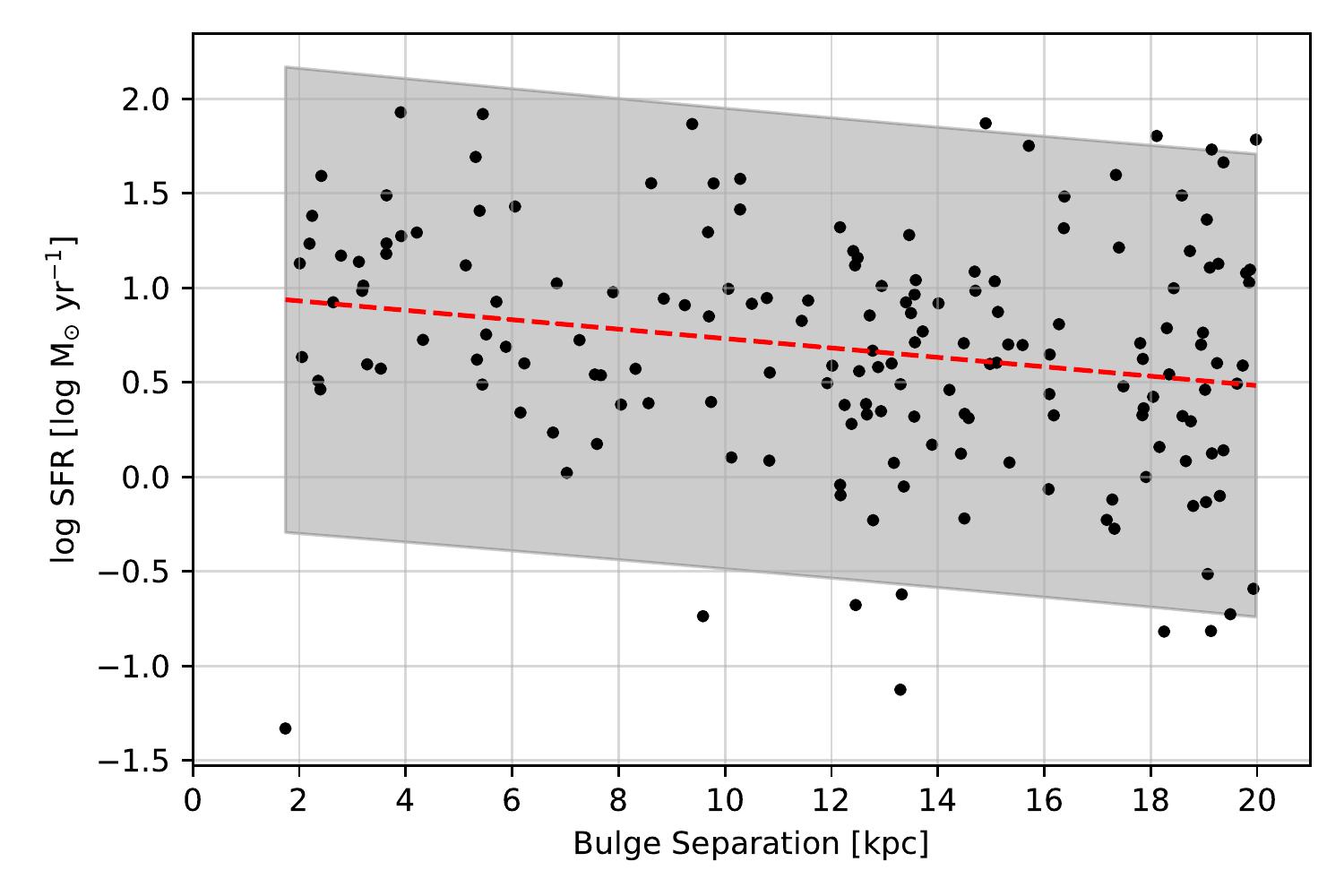}
	\vspace{-6pt}
	\includegraphics[width=0.48\textwidth]{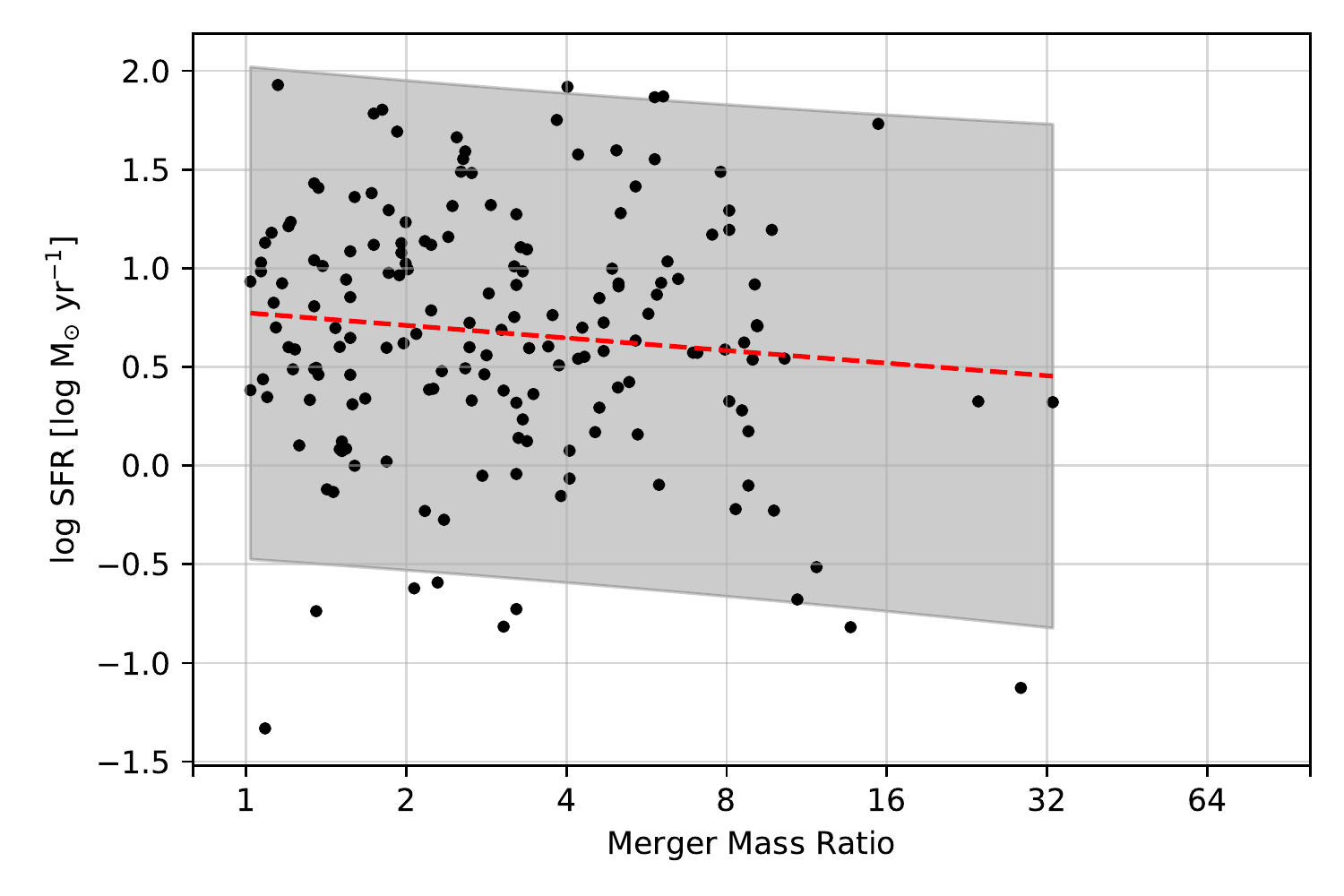}
	
	\caption{Relations between AGN and host galaxy properties with the merger parameters of bulge separation (left) and merger mass ratio (right); from top to bottom, the AGN and host galaxy properties examined are: nuclear column density, AGN luminosity, and SFR. The overlaid red dashed lines and gray regions are linear fits to the data and the fits' 95\% confidence regions, respectively. None of the linear fits show any significant correlations except for SFR as a function of separation; this is due to correlations with galaxy mass (see Section \ref{subsec: props vs merge param} and Figure \ref{plot: gal props 3}). Note that the scale for the merger mass ratio axis is Log$_{2}$.}
	\label{plot: gal props 2}
	\vspace{6pt}
\end{figure*}

\subsection{Comparison to the General AGN Sample}\label{subsec: acs-agn compare}

With this being the largest catalog of offset and dual AGNs yet assembled, we examine how this population behaves and relates to more general populations, such as the population of all AGN host galaxies.

At first glance, it seems that our merger subsample is over-represented at lower redshifts (and under-represented at higher redshifts) when compared with the general AGN population of the ACS-AGN, as seen in Figure \ref{plot: z plots}. Specifically, we find that our offset and dual AGN distribution increases until $z \sim 1$ and then decreases. This is surprising given that the merger fraction is thought to increase with redshift, peaking at redshifts near $z=1.5$ \citep[e.g.][]{Carlberg1990}, with work studying deep surveys confirming those predictions \citep[e.g.][]{Ryan2008, Stott2013}. However, our distribution is in agreement with the merger fraction for massive galaxies in the UltraVISTA/COSMOS catalog, as shown in \cite{Man2016}. They explain that the reason for the drop in merger fraction above $z=1$ in their sample is due to being incomplete at higher redshifts as low surface brightness partners become undetectable. 

Therefore, we test whether the trends seen in our offset and dual AGNs' redshift distribution are also due to this selection bias. In order to do this, we adjusted the surface brightnesses and pixel separations of our sample to what would be observed if they were at the highest redshift in our sample, $z=2.5$; specifically, for surface brightness we multiplied by a factor of $(d_{L}(z)/d_{L}(z=2.5))^2$, where $d_{L}(z)$ is the luminosity distance at the given redshift, $z$; this follows the known relation of surface brightness being proportional to $(1+z)^{-4}$ \citep[e.g.][]{Tolman1930, Hubble1935, Sandage1961}. We found that the bulge separations of our offset and dual AGNs were not a limiting factor, with all being able to be detected even when adjusted to a $z=2.5$ frame. However, only 16 out of the 220 dual and offset AGNs in our sample had sufficient surface brightness to be able to be detected using the methods outlined in Section \ref{sec:Bulge Select} when adjusted to a $z=2.5$ frame. 

Therefore, we are heavily biased towards selecting only the brightest bulges at high redshifts. If we only examine the subsample of offset and dual AGNs that could be observed even when adjusted to the $z=2.5$ frame (as seen in Figure \ref{plot: z plots}), we find that our distribution peaks near $z=1.5$, falling off at low redshift due to the decrease in survey volume observed as redshift decreases and dropping off rapidly at high redshift due to surface brightness dimming and survey depth limits. Therefore, we find that the redshift distribution of our offset and dual AGN sample recovers the expected merger fraction distribution with redshift of previous works \citep[e.g.][]{Carlberg1990,Ryan2008,Stott2013}, once this has been corrected for.

The distributions of the rest of the AGN and host galaxy properties ($M_{*}$, $L_{AGN}$, SFR, and $N_{H}$) for the offset and dual AGNs closely mirror that of the general AGN population of the ACS-AGN Catalog. This can be seen in Figures \ref{plot: z plots} \& \ref{plot: gal props 1}, and is verified by Kolmogorov-Smirnov (KS) tests for each. 

We also examined the AGN and host galaxy properties of our offset and dual AGNs in comparison to samples of non-merging AGN host galaxies matched by redshift and galaxy stellar mass. To do this, we matched each offset and dual AGN in our sample to a set of non-merging AGN host galaxies from the ACS-AGN catalog that had redshifts within 0.1 and galaxy stellar masses within 0.25 dex of the offset or dual AGN system. We required that each dual and offset AGN had a matched sample of at least 10 systems in order to be analyzed. In total, we examined 163 of the 220 dual and offset AGNs along with their matched samples, with the minimum, median, and maximum number of ACS-AGN galaxies in the  matched samples being 10, 60, and 133, respectively. We then calculated how the offset and dual AGNs' $L_{AGN}$, SFR, and $N_{H}$ values differed from their matched non-merging sample's mean $L_{AGN}$, SFR, and $N_{H}$ values. The distribution of the differences between the offset and dual AGNs and their matched samples $L_{AGN}$, SFR, and $N_{H}$ values are shown in Figure \ref{plot: gal props 1}. We find that the means of the difference distributions of these properties are all approximately zero, and therefore that the $L_{AGN}$, SFR, and $N_{H}$ values of the offset and dual AGNs do not significantly differ from that of non-merging AGN host galaxies, even when matched in redshift and galaxy stellar mass.

From these results, we specifically find that at high AGN luminosities, offset and dual AGNs are not overrepresented compared to the general AGN population of the ACS-AGN. Therefore we find that mergers do not preferentially trigger the most luminous AGN, but instead host AGN with similar luminosities to AGN not in mergers; this is true not only when examining the total sample, but also for the minor mergers (merger mass ratio $>4$) and major mergers (merger mass ratio $<4$) separately. This is in disagreement with theoretical predictions made by \cite{Hopkins2009} and observations by \cite{Treister2012}, but is in agreement with observational work by \cite{Kocevski2012} and \cite{Villforth2014}.

 We also do not find any evidence that SFR is enhanced in offset and dual AGNs, instead finding that offset and dual AGNs do not exhibit any shift towards higher SFRs when compared to the overall active galaxy population. Other works have found that star formation is enhanced in mergers in general \citep[e.g.][]{Joseph1985, Knapen2015}. Our results indicate that the presence of offset and dual AGNs makes these merging systems unique in this regard. The lack of a shift to higher SFRs in offset and dual AGNs may indicate that the presence of an AGN in the merging system inhibits the increase in SFR seen in non-AGN host mergers. If this is so, then this is a key piece of observational evidence pointing toward the ability of AGNs to impart negative feedback during galaxy mergers and shut down star formation in their host galaxies.
 
 Lastly, we find that the nuclear column densities of offset and dual AGNs are similar on average to those of the overall active galaxy population and that there is no enhancement present in offset and dual AGNs even when matched in redshift and galaxy stellar mass. However, the distribution is highly asymmetric, showing a longer tail of less obscured systems out to -2 dex, while the distribution peaks at the moderately increased obscuration value of +0.5 dex.

\begin{figure*}
	\centering
	\includegraphics[width=0.48\textwidth]{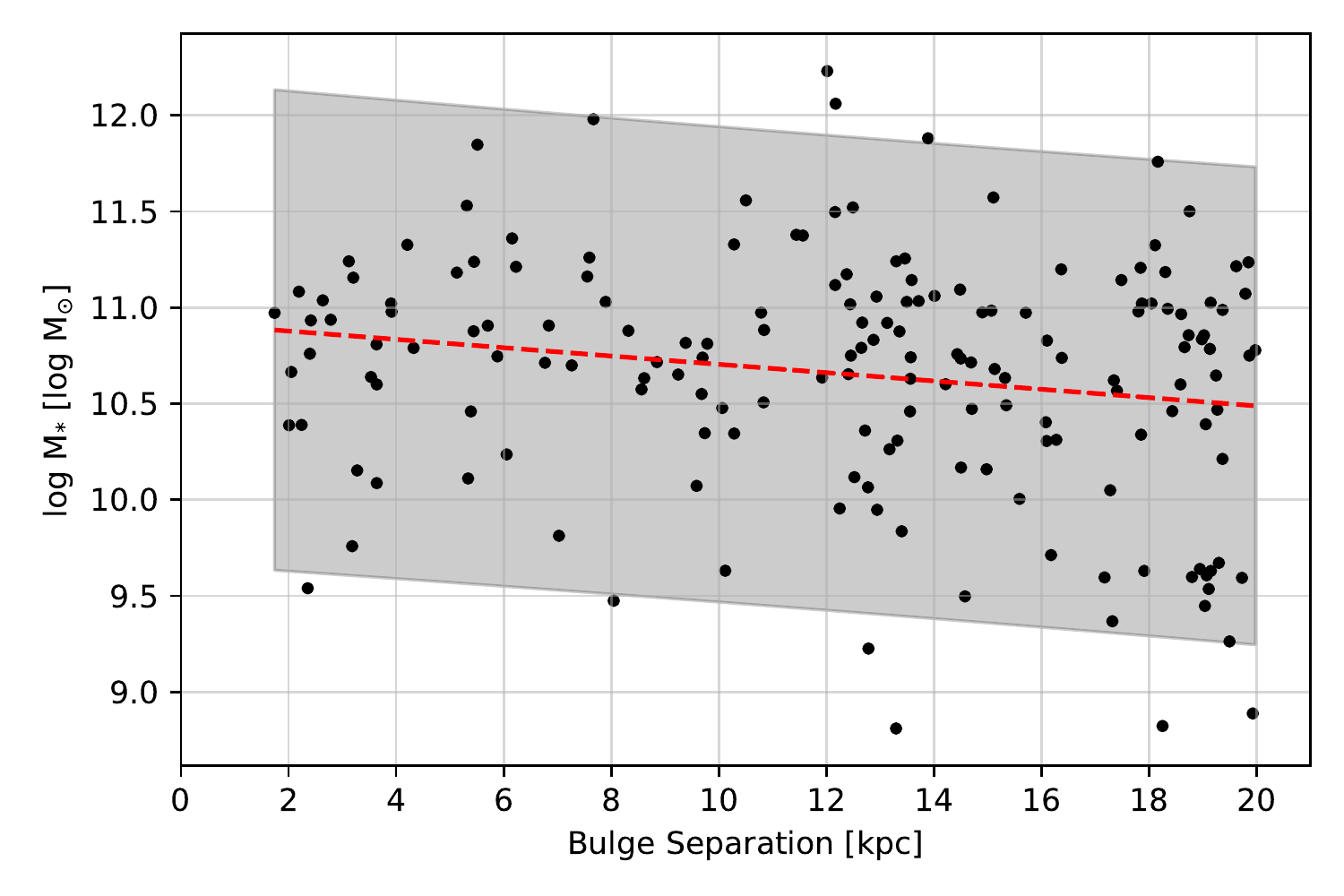}
	\vspace{-6pt}
	\includegraphics[width=0.48\textwidth]{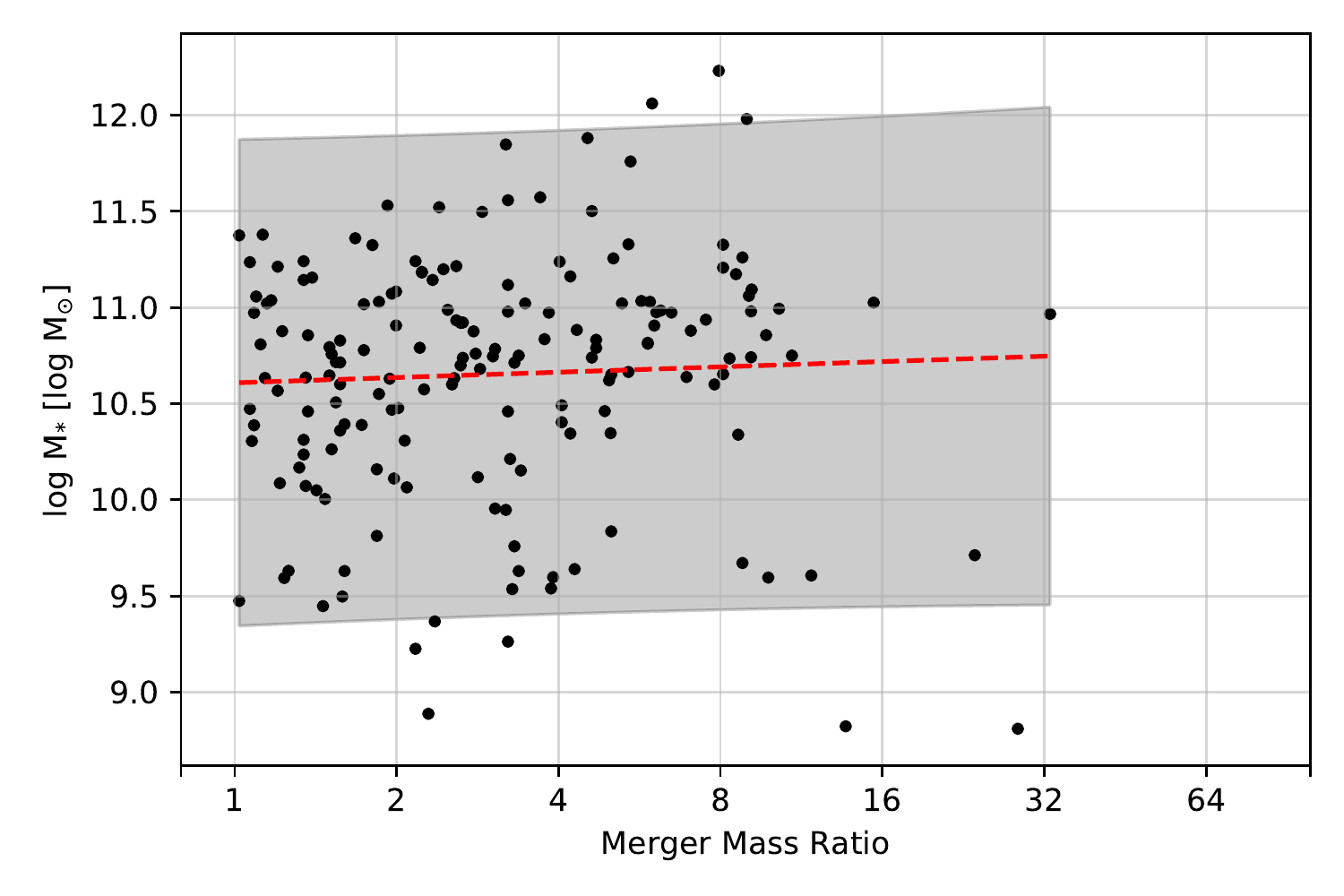}
	\includegraphics[width=0.48\textwidth]{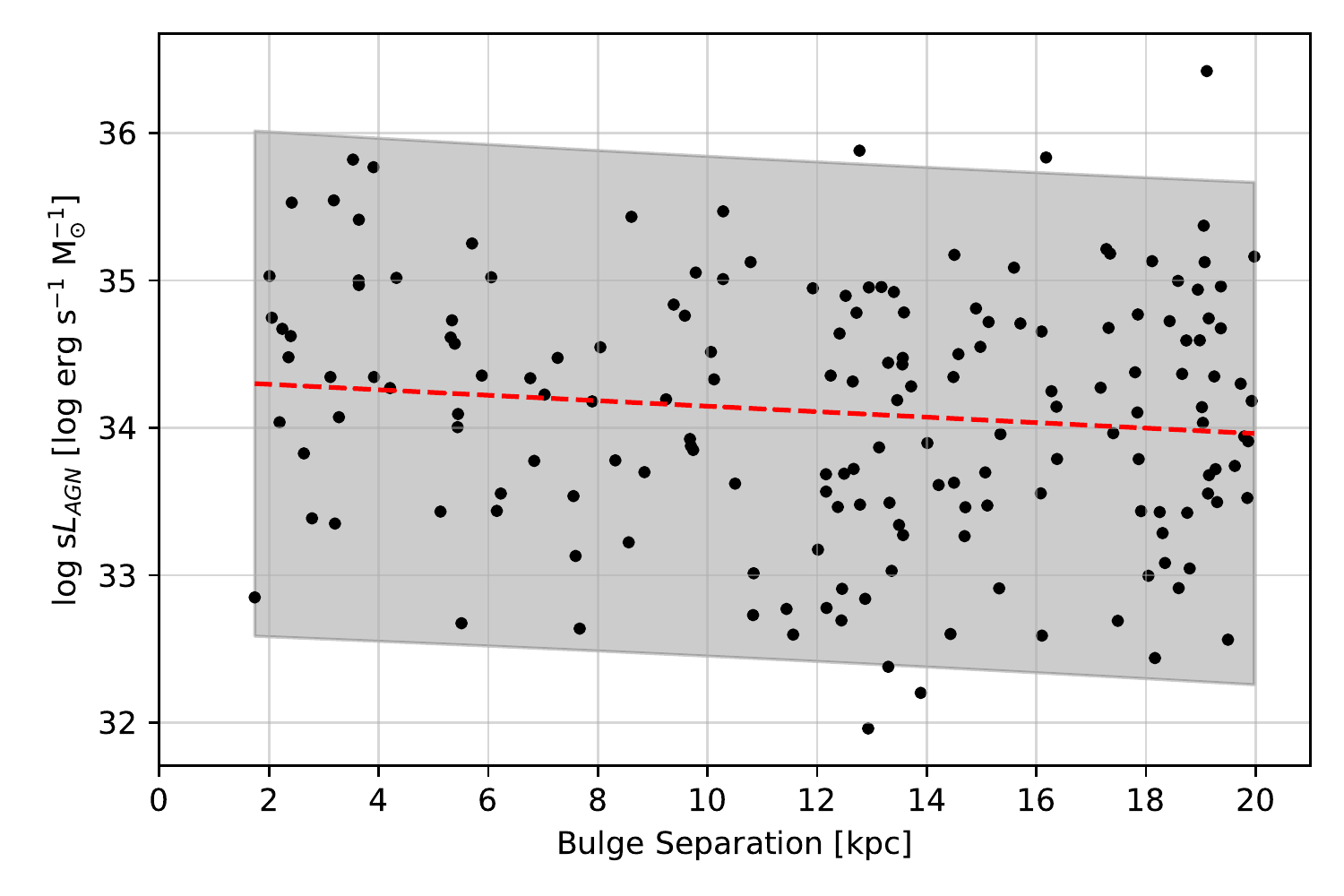}
	\vspace{-6pt}
	\includegraphics[width=0.48\textwidth]{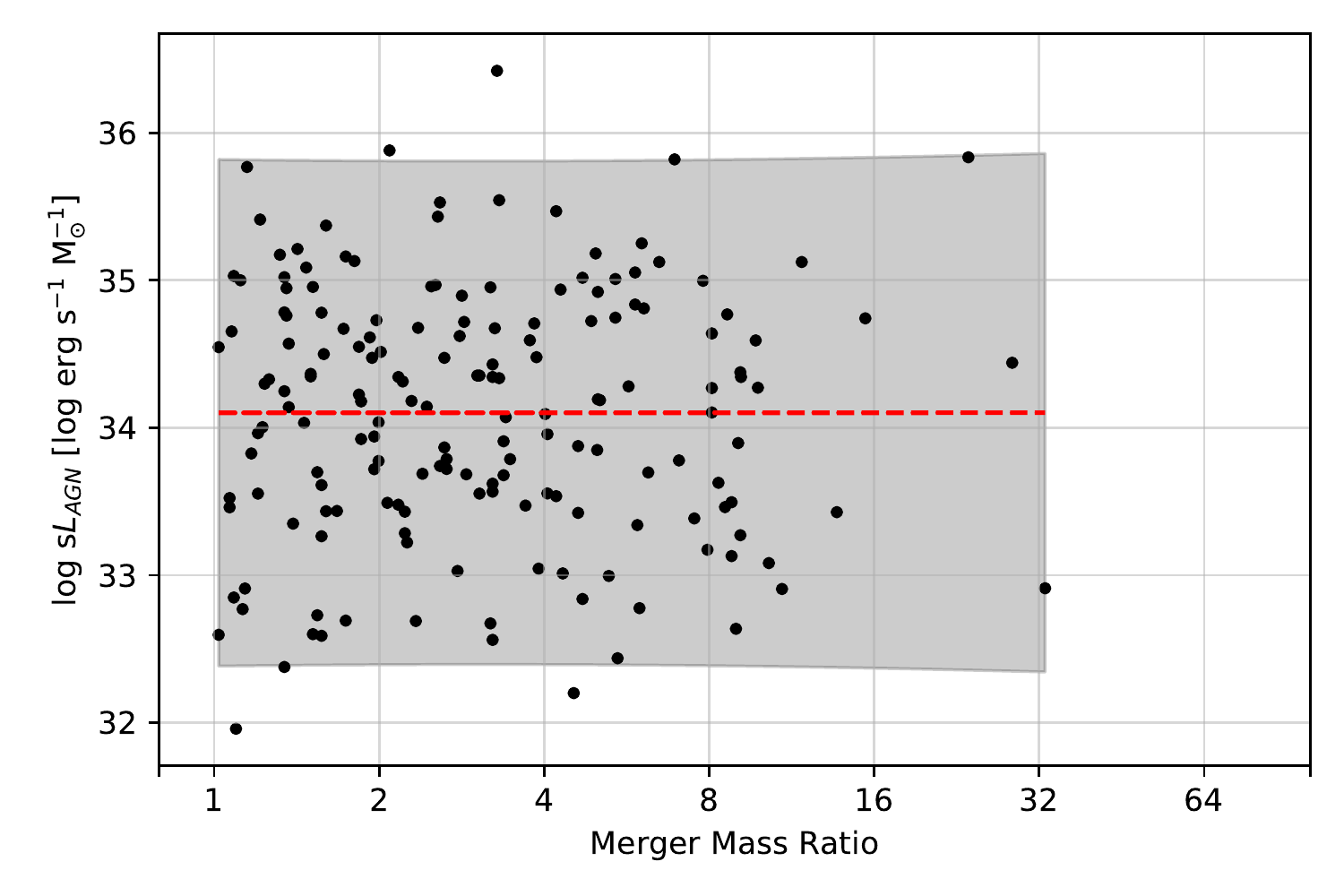}
	\includegraphics[width=0.48\textwidth]{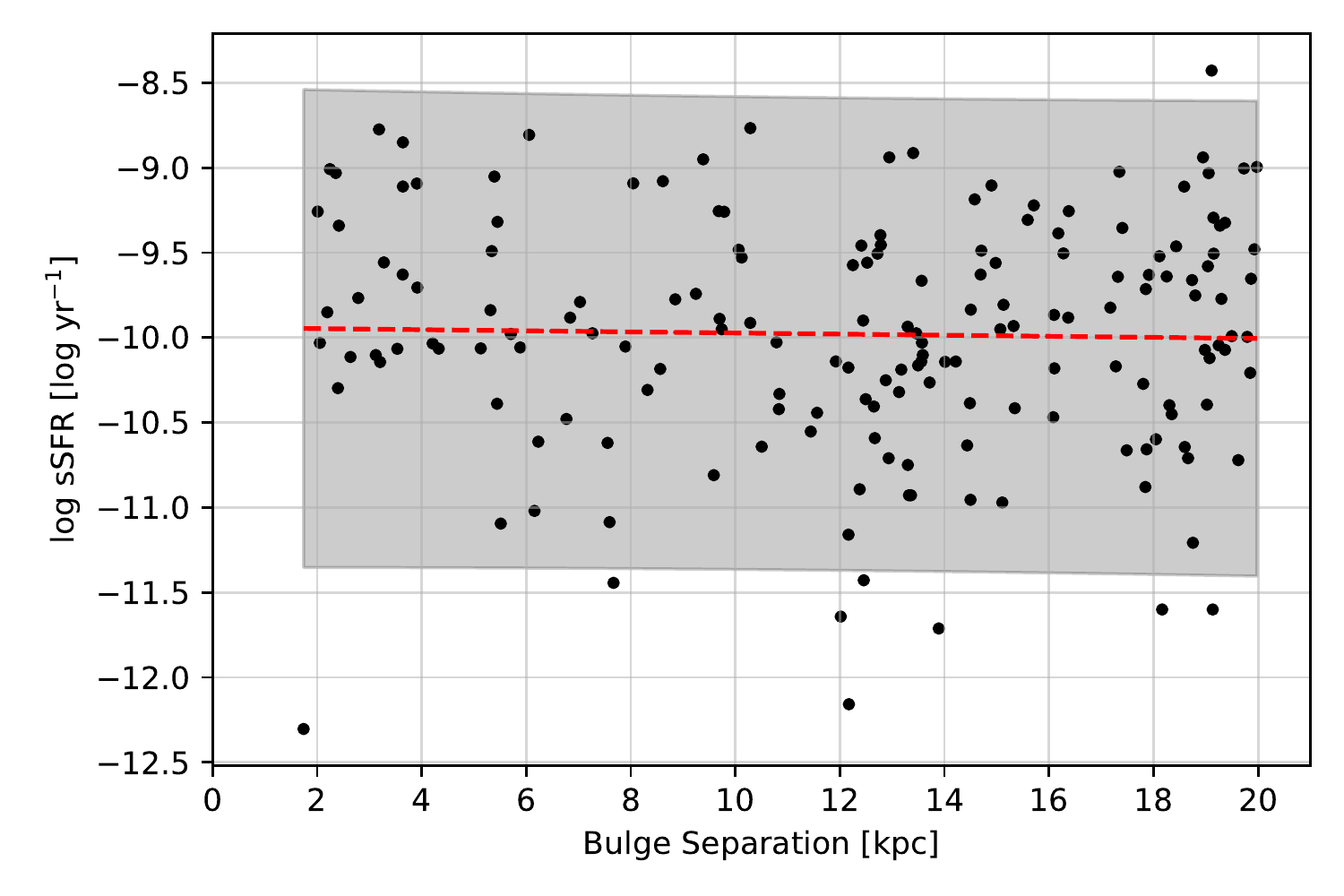}
	\includegraphics[width=0.48\textwidth]{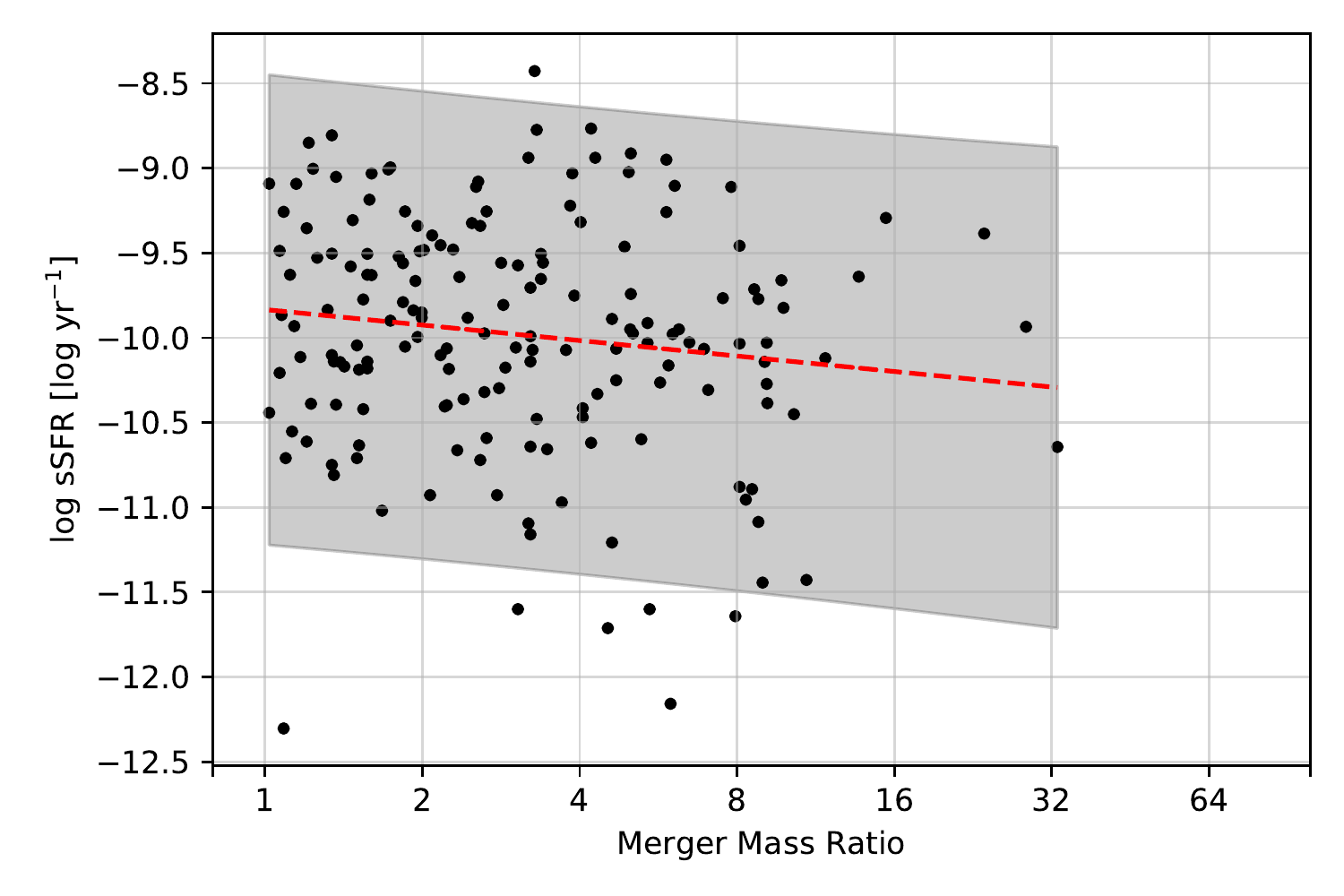}
	\vspace{-12pt}
	\caption{Relations between specific SFR and specific AGN luminosity ($L_{AGN}/M_{*}$)with the merger parameters of bulge separation (left) and merger mass ratio (right). The overlaid red dashed lines and gray regions are linear fits to the data and the fits' 95\% confidence regions, respectively. We see that by accounting for known correlations with galaxy mass, neither specific SFR or specific AGN luminosity show any significant correlations with bulge separation or merger mass ratio.}
	\label{plot: gal props 3}
	\vspace{6pt}
\end{figure*}

\vfill
\subsection{AGN and Host Galaxy Properties\\Are Not Correlated with Merger Parameters}\label{subsec: props vs merge param}

Lastly, we examine the AGN and host galaxy properties of offset and dual AGNs as a function of their merger properties. We specifically examine relations between nuclear column density, AGN luminosity, and SFR to search for correlations in our data. 

Merger events should drive significant material inwards towards the center of the merging system; this material can fuel star formation and SMBH growth \citep[e.g.][]{Hopkins2008,Hopkins2009}. There has been a substantial amount of work done predicting and searching for evidence of increased star formation, and increased SMBH growth in mergers. Star formation is expected to increase as separation decreases in merging systems \citep[e.g.][]{Ellison2008,Patton2013a}, as well as increase as merger ratio approaches unity \citep[e.g.][]{Somerville2001,Cox2008,Ellison2008}. Since SMBHs are thought to be fed by the same material that is enhancing star formation in these scenarios, it is also predicted that AGN triggering increases as separation decreases to small separations \citep[e.g.][]{Ellison2011,VanWassenhove2012,Koss2012,Blecha2013,Satyapal2014}; whether the most luminous AGN are found in mergers with mass ratios near unity is debated \citep[e.g.][]{Kocevski2012,Treister2012,Villforth2014,Villforth2017}. If true, this increase would follow from the expected increase in available material near the center of the merging system; this has been hinted, with observations suggesting that obscured AGN are more likely to reside in mergers \cite{Kocevski2015, Ricci2017, Donley2018, Pfeifle2019}, but the sample sizes have been limited in these studies.

When looking at our relations shown in Figure \ref{plot: gal props 2}, it is apparent that there is a large amount of scatter in the data, with no tight correlations emerging. Overlaid on top of the data are linear fits; these correspond to exponential and logarithmic fits when examining these properties as functions of bulge separation and merger mass ratio, respectively. Excluding SFR as a function of bulge separation, we find that none of the fits have slopes that are significantly non-zero (p-value $<$ 0.05) and that no pair is significantly correlated, as determined by its Spearman rank-order correlation coefficient (p-value $<$ 0.05).

However, we know that the ACS-AGN galaxies show correlations between galaxy mass and SFR as well as galaxy mass and AGN luminosity (see S20 for more detail); this relation has also been shown to exist for galaxies in mergers \citep{Barrows2017b}. As we can see in Figure \ref{plot: gal props 3}, galaxy mass is significantly inversely correlated with separation in our sample; this is verified by examining the significance of its fit and Spearman rank-order correlation coefficient. Therefore we examine the specific SFR (sSFR) and AGN luminosity divided by galaxy stellar mass ($sL_{AGN}$), in order to correct for any galaxy mass -- separation correlation and galaxy mass -- merger ratio correlation.

While SFR is significantly correlated with bulge separation, we find that this is due to the correlation between galaxy mass and separation. When examining sSFR and specific AGN luminosity, we find that neither have significant linear trends nor are they correlated at significant levels to bulge separation or merger mass ratio. This can be seen in Figure \ref{plot: gal props 3}. Where \cite{Barrows2018} find that AGN luminosity increases as merger mass ratio decreases, this work does not find such a relationship.

From these data, we find that there is no observational evidence that nuclear column density, AGN luminosity, or SFR is correlated with bulge separation or merger mass ratio for offset and dual AGNs. This stands in contrast to many of the findings outlined above related to SFR increasing as separation decreases and as merger ratio approaches unity in the general population of mergers. Since the primary difference between these two populations is the presence of AGNs, this finding provides another piece of evidence supporting the hypothesis that an AGN can impart negative feedback on its host galaxy, maintaining a lower level of star formation than would be expected in a merger without AGN. We also find that the most luminous AGN are not preferentially triggered by major mergers; this is in agreement with previous findings by \cite{Kocevski2012} and \cite{Villforth2014,Villforth2017}. 

\begin{figure*}[!]
	\centering
	\includegraphics[width=0.49\textwidth]{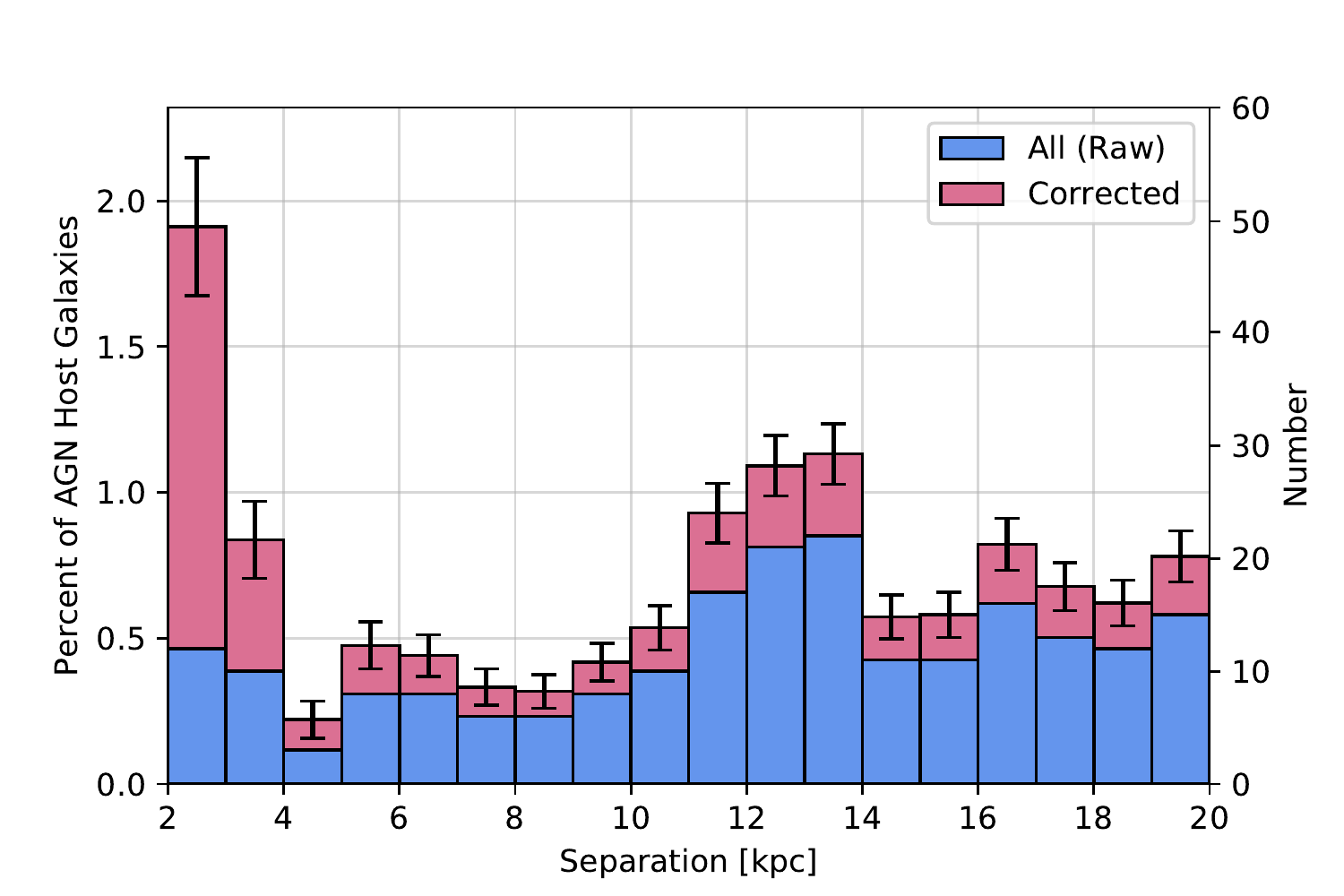}
	\vspace{-4pt}
	\\
	\includegraphics[width=0.49\textwidth]{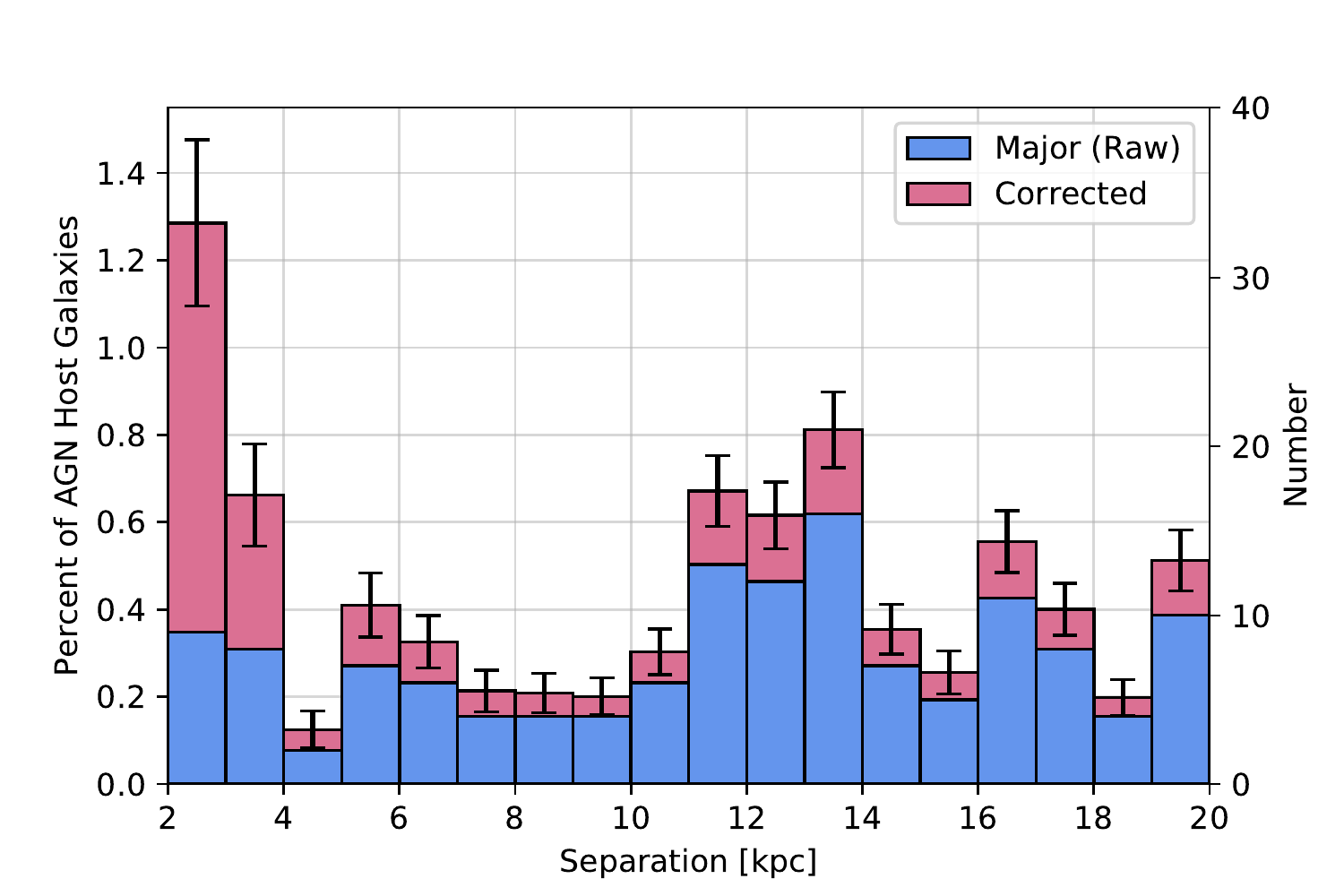}
	\includegraphics[width=0.49\textwidth]{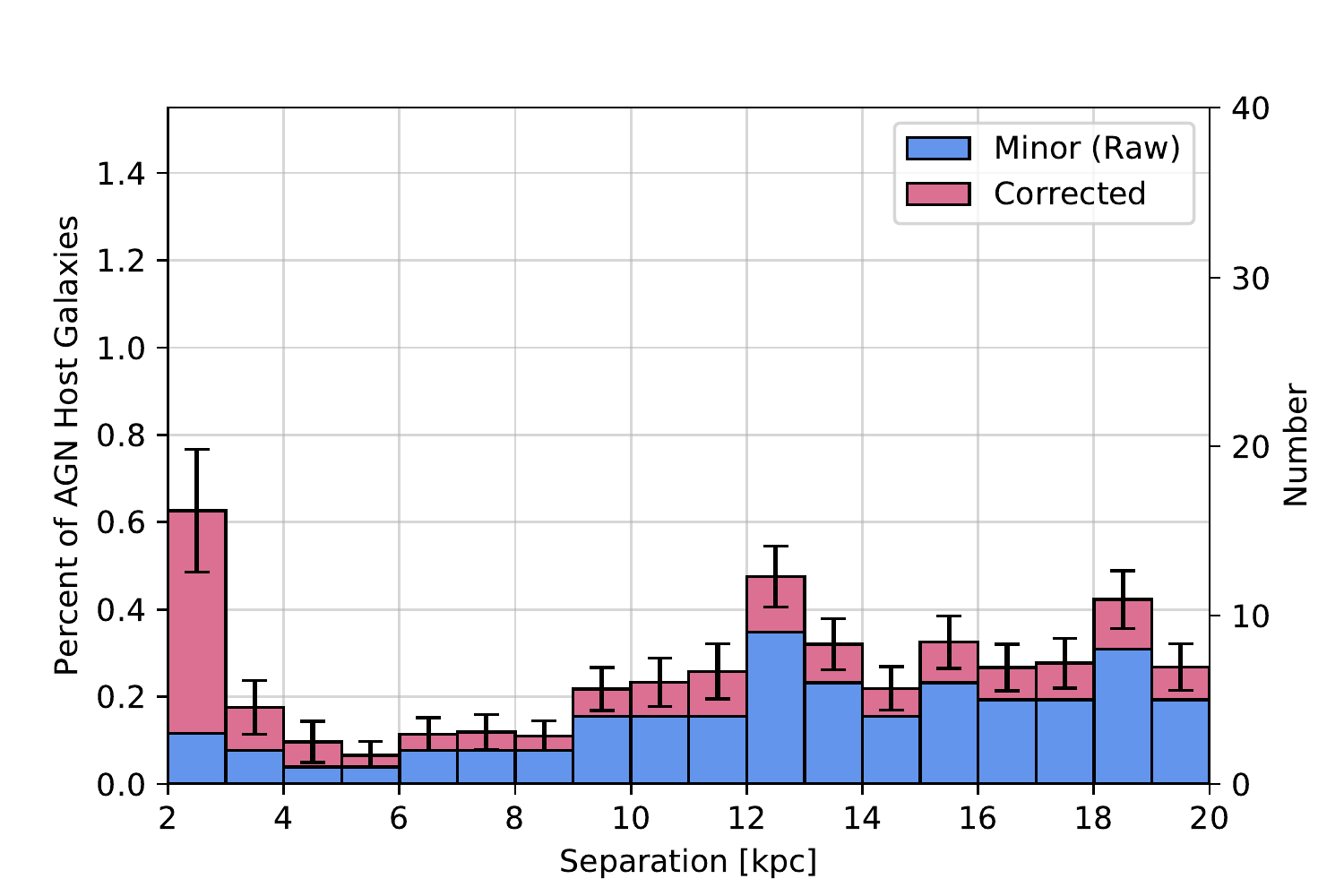}
	\caption{Histogram of percentage of AGNs in mergers as a function of separation; raw numbers are reported as well as bias corrected values. The top plot shows all of our sample, while the bottom left shows only major mergers (ratio $<$ 4) and the bottom right shows only minor mergers (ratio $>$ 4). Note that AGN activation peaks at separations less than 4 kpc and that there is a bump, more apparent in the major merger subsample than minor merger subsample, near separations 11 -- 14 kpc.}
	\label{plot: sep histogram}
\end{figure*}

\section{Results} \label{sec: results}
\subsection{AGN Activation Peaks at the \\ Smallest Bulge Separations}\label{subsec: result AGN frac}

One of the major open questions that this sample of offset and dual AGNs can address is whether the fraction of AGNs in mergers peaks at separations below 10 kpc. Theoretical work predicts that SMBH growth peaks at separations of either 1 -- 10 kpc \citep{VanWassenhove2012} or 0.1 -- 2 kpc \citep{Blecha2013}. While observations have found that AGN fraction increases from 100 -- 10 kpc, the trend could not be constrained below 10 kpc due to the limited samples at small separations \citep{Ellison2011, Koss2012, Barrows2017a}. 

We find that AGN activation increases significantly below 10 kpc. In Figure \ref{plot: sep histogram}, we see that the AGN activation is mostly flat from 20 -- 14 kpc, has a bump from 14 -- 11 kpc, drops slightly from 11 -- 4 kpc, and then increases significantly and peaks at our smallest separation bin, 3 -- 2 kpc. This distribution is in qualitative agreement with previous work examining simulations of dual AGNs by \cite{Rosas-Guevara2019} that shows a bump in dual AGN fraction near 14 kpc and a rise as separations near 5 kpc. It should be noted that we had one system observed between 1.5 and 2 kpc, which is not included in this analysis because it is the only system in that bin.

We also analyze both the major and minor merger sub-populations of our sample in Figure \ref{plot: sep histogram}; we define major mergers as those with merger mass ratios $<$4 and minor mergers as those with merger mass ratios $>$4. We see similar distributions as the total sample, with the exception that major mergers see a more significant bump from 14 -- 11 kpc --- both larger and wider --- than the bump seen with minor mergers. These bumps coincide with separations corresponding to the merger's first pericenter passage, typically seen at separations from 10 --  20 kpc in simulations, when significant gas would be driven inwards, triggering AGN activation \citep{VanWassenhove2012, Blecha2013, Rosas-Guevara2019}. This also explains why the bump is more significant in major mergers versus minor mergers, as first pericenter passage in major mergers should be more dynamic, driving more gas inwards than minor mergers, and more readily triggering AGN activation.

Therefore, we confirm the theoretical predictions of \cite{VanWassenhove2012} and \cite{Blecha2013}, and are able to extend the trends seen by \cite{Ellison2011}, \cite{Koss2012}, and \cite{Barrows2017a}. Further, we find that we are in good agreement with the smallest separation bin of \cite{Koss2012}, which studied a small sample of moderate-luminosity, ultra-hard X-ray selected AGN at low redshifts. They find that 7.8$\pm 1.8\%$ of AGN are found in mergers at separations $<15$ kpc, while we find that 9.3$\pm 1.3\%$ of AGN are found in mergers at those separations, after accounting for biases.

It is uncertain whether AGN activation and SMBH growth continues to increase below 2 kpc. As was discussed in Section \ref{subsec: fitting}, a reasonable estimate of the physical size of a stellar bulge is approximately 1.5 kpc. Further, 1.5 kpc is very close to the limit at which we can accurately resolve two bulges, if they are even physically small enough to exist discretely at those separations. Therefore, we cannot reliably use offset and dual AGN discovered by selecting multiple stellar bulges to study AGN activation and SMBH growth below 2 kpc.

\subsection{AGN are Preferentially \\ Found in Major Mergers}\label{subsec: result AGN major}

The second primary question that can be addressed with our sample of offset and dual AGNs is whether merger mass ratio affects AGN activation. Many others have examined whether major mergers trigger the most luminous AGN. As previously discussed, theoretical work by \cite{Hopkins2009} predicts that major mergers trigger the most luminous AGN, while observational studies on the subject are not in agreement, with some in finding they do \citep[e.g.][]{Treister2012}, and some finding no relation \citep[e.g.][]{Kocevski2012,Villforth2014,Villforth2017}. 

\begin{figure}[!t]
	\begin{center}
		\includegraphics[width=0.47\textwidth]{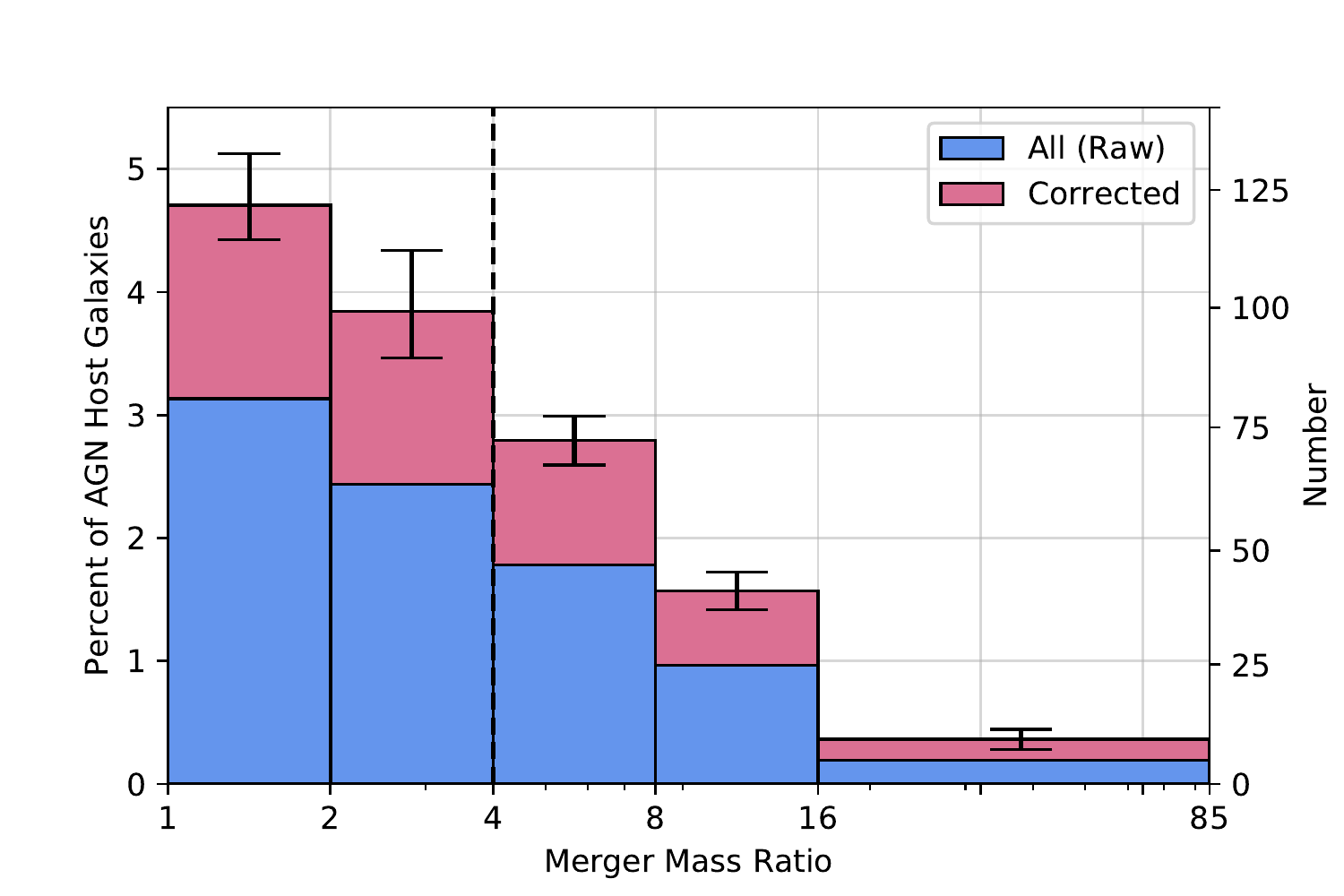}
	\end{center}
	\vspace{-12pt}
	\caption{Histogram of percentage of AGNs in mergers as a function of merger mass ratio; raw numbers are reported in addition to the bias corrected values. Note that the scale for the merger mass ratio axis is Log$_{2}$, and that the last bin spans merger mass ratio values from 16 to 85 due to the small number of galaxies in that range.. The dashed vertical line at merger mass ratio $=$ 4 separates major mergers to the left and minor mergers to the right. Note that AGNs in mergers at separations $<20$ kpc are preferentially found in major mergers and that the percent of AGNs decreases as ratio increases as a log base 2 function with slope $\approx-1$ from merger mass ratio values of 1 to 16.}\label{plot: ratio histogram}
\end{figure}

While we also find that there is no relation between merger mass ratio and AGN luminosity (as discussed in Section \ref{subsec: props vs merge param}), we do find that AGNs are preferentially found in major mergers; this can be seen in Figure \ref{plot: ratio histogram}. Using our corrected values and a merger mass ratio cutoff value of 4:1 between major and minor mergers, we find that 8.5$\pm$0.9\% of all AGNs are found in major mergers at separations $<$20 kpc and 4.7$\pm$0.4\% of all AGNs are found in minor mergers (with merger mass ratios greater than 4 and less than 85) at separations $<$20 kpc. In total we find that $13.3\pm1.7$\% of all AGNs are found in mergers at separations $<$20 kpc. The overabundance of AGN in major mergers compared to minor mergers is especially significant because it is thought that minor mergers outnumber major mergers three to one \citep[e.g.][]{Bertone2009,Lotz2011}; this means that AGN are overrepresented in major mergers by approximately a factor of six when compared to what would be expected if AGN activation had no dependence on merger mass ratio.

We also find that the fraction of AGNs in mergers at separations $<20$ kpc follows a logarithmic relationship from ratios of 1 to 16 ---  we find a relationship of: percent of AGN in merger mass ratio bin $=-$Log$_{2}$(MassRatio)$+5$, where MassRatio is the center value of the bin. Note that due to the nature of a histogram, this specific relation only applies when using bin widths of powers of two, but a logarithmic relation would remain regardless.

Last, we can compare our fractions of AGN activation in the lowest merger mass ratio bins to that found by \cite{Koss2012}. They find that the fraction of their AGNs in each bin is 4.2$\pm$1\%, 1.8$\pm$0.6\%, and 1.8$\pm$0.6\% for merger mass ratio bins from 1 -- 2, 2 -- 4, and 4 -- 8, respectively; they did not have any systems with merger mass ratios greater than 8. In comparison we find that the fraction of AGNs in each bin is 4.7$\pm$0.4\%, 3.8$\pm$0.5\%, and 2.8$\pm$0.2\%, respectively. Comparing these values, we find that while we are in good agreement for mergers with mass ratios nearest unity, our approach finds more offset and dual AGNs at higher merger mass ratio values than the \cite{Koss2012} sample. We note that the fraction of AGN that we find to be in mergers is significantly higher than values predicted by \cite{Yu2011} and \cite{Rosas-Guevara2019}; this is most likely due to those works focusing on dual AGN specifically, while this work includes offset AGN as well.

\pagebreak
\section{Conclusions} \label{sec: conclusions}

Here we have detailed the identification and analysis of 220 offset and dual AGNs observed with \textit{HST}/ACS at $0.2<z<2.5$ and with stellar bulge separations $<$20 kpc. These offset and dual AGNs are a subsample of the AGN host galaxies in the ACS-AGN Catalog and were selected as AGN using mid-infrared and X-ray selection criteria applied to \textit{Spitzer} and \textit{Chandra} data. This sample and their associated AGN, AGN host galaxy, and merger properties are made available in the ACS-AGN Merger Catalog. We have analyzed this sample, with our main findings summarized below.

\begin{enumerate}
	\item We find that AGN activation increases significantly at bulge separations from 4 to 2 kpc. We also note a bump at separations from 14 to 11 kpc that is more significant in major mergers than minor mergers; this is likely attributable to first pericenter passage occurring near these separations. 
	\item We find that AGNs in mergers at separations $<$20 kpc are preferentially found in major mergers. Specifically, using values corrected for selection effects related to bulge separation and merger mass ratio, we find that 8.5$\pm$0.9\% of all AGNs are found in major mergers (merger mass ratios from 1 to 4) and 4.7$\pm$0.4\% of all AGNs are found in minor mergers (merger mass ratios from 4 to 85), bringing the total percent of AGNs found in mergers to 13.3$\pm$1.7\%. We also find that the percentage of AGNs found at different merger mass ratios follows a logarithmic relation, decreasing as merger mass ratio increases.
	\item We find that mergers do not trigger the brightest AGNs, but instead mergers at separations $<$20 kpc trigger AGNs with a similar distribution of luminosities as that of the general AGN population. This is true of all mergers, including the major and minor merger sub-populations.
	\item We find that SFR, AGN luminosity, and nuclear column density have no significant correlations or dependencies on bulge separation or merger mass ratio, once known correlations with galaxy mass have been accounted for. Further, we find that the distributions of these values for AGNs and AGN host galaxies in mergers at separations $<$20 kpc do not significantly differ from the distributions of AGNs and AGN host galaxies in general.
\end{enumerate}

These findings show that in mergers where the stellar bulges are separated by $<$20 kpc, bulge separation and merger mass ratio play an important role in the activation of AGNs, but do not significantly enhance star formation or the luminosity of AGNs in galaxy mergers. This implies that there may be AGN feedback involved in these systems that heats and/or blows out excess material, inhibiting AGN growth and slowing star formation rapidly after AGN activation, returning the system to AGN luminosity and star formation rates typically seen in non-mergers.

\section*{Acknowledgments}
Support for this work was provided by NASA's Astrophysics Data Analysis program, grant number NNX15AI69G, the CU Boulder / JPL Strategic University Research Partnership program, the National Science Foundation's Graduate Research Fellowship program, and a Chance Irick Cooke Fellowship. The work of DS was carried out at the Jet Propulsion Laboratory, California Institute of Technology, under a contract with NASA. RJA was supported by FONDECYT grant number 1191124. This work utilized the RMACC Summit supercomputer, which is supported by the National Science Foundation (awards ACI-1532235 and ACI-1532236), the University of Colorado Boulder, and Colorado State University. The Summit supercomputer is a joint effort of the University of Colorado Boulder and Colorado State University. The authors would also like to thank Rebecca Nevin and Hayley Roberts for providing support to this work. 

\bibliographystyle{apj}
\bibliography{../../library}

\end{document}